\documentclass[a4paper,11pt]{article} 
\usepackage [latin1]{inputenc}
\voffset - 2truecm
\hoffset - 1.4 truecm
\def\beq{\begin{equation}}
\def\eeq{\end{equation}}
\usepackage{amsmath,amsfonts,amssymb,amsthm}   
\usepackage{mathrsfs}
\usepackage{multicol}
\usepackage{fancybox} 
\usepackage{array}
\usepackage{color}
\usepackage{graphicx}
\begin{document}
\def\Sym{{\rm Sym}}
\def\R{\mathbb R}
\def\N{\mathbb N}
\def\C{\mathbb C}
\def\Z{\mathbb Z}
\def\Q{\mathbb Q}
\def\bigat{\mbox {\Large (}}
\def\bigct{\mbox {\Large )}}
\def\Ker{{\cal N}}
\def\dom{{\rm dom}\;}
\def\supp{{\rm supp}\;}
\def\pf{\rm pf}
\def\arccot{{\rm arccot}}
\def\nequiv{\equiv \!\!\!\!\! / \,\, }
\def\tr{{\rm tr}\,}
\def\sgn{{\rm sgn}\,}
\def\vol{{\rm vol}}
\def\sym{{\rm sym}}
\def\skw{{\rm skw}}
\def\sp{{\rm sp}}
\def\dual{{\rm dual}}
\def\ctg{{\rm cot}}
\def\cot{{\rm cot}}
\def\sec{{\rm sec}}
\def\csc{{\rm csc}}
\def\diag{{\rm diag}}
\def\arcos{{\rm arccos}}
\def\bigaq{\mbox {\Large [}}
\def\bigcq{\mbox {\Large ]}}
\def\cA{{\mathcal A}}
\def\cB{{\mathcal B}}
\def\cC{{\mathcal C}}
\def\cD{{\mathcal D}}
\def\cE{{\mathcal E}}
\def\cF{{\mathcal F}}
\def\cG{{\mathcal G}}
\def\cH{{\mathcal H}}
\def\cX{{\mathcal X}}
\def\cI{{\cal I}}
\def\cL{{\mathcal L}}
\def\cN{{\cal N}}
\def\cO{{\cal O}}
\def\cR{{\mathcal R}}
\def\cP{{\cal P}}
\def\cS{{\cal S}}
\def\cT{{\cal T}}
\def\cU{{\cal U}}
\def\cV{{\cal V}}
\def\a{[\!\!\,[}
\def\c{]\!\!\,]}
\def\one{1\!\!\,{\rm l}}
\def\m{{\scriptscriptstyle -}}
\def\p{{\scriptscriptstyle +}}
\def\pp{{\scriptscriptstyle {++}}}
\def\sfot{{\scriptstyle{\frac 1 2}}}
\def\sfoth{{\scriptstyle{\frac 1 3}}}
\def\sfof{{\scriptstyle{\frac 1 4}}}
\def\sftth{{\scriptstyle{\frac 2 3}}}
\def\crux{\,{\times \!\!\!\!\! \times}\,}
\def\lbox{\mbox{\raisebox{-.25ex}{\Large$\Box$}}}
\def\boldeta{\mbox{\boldmath $\eta$}}
\def\bepsilon{\mbox{\boldmath $\epsilon$}}
\def\bgamma{\mbox{\boldmath $\gamma$}}
\def\bomega{\mbox{\boldmath $\omega$}}
\def\btheta{\mbox{\boldmath $\theta$}}
\def\bphi{\mbox{\boldmath $\phi$}}
\def\bpsi{\mbox{\boldmath $\psi$}}
\def\bxi{\mbox{\boldmath $\xi$}}
\def\bvarphi{\mbox{\boldmath $\varphi$}}
\def\bvarpi{\mbox{\boldmath $\varpi$}}
\def\bchi{\mbox{\boldmath $\chi$}}
\def\btau{\mbox{\boldmath $\tau$}}
\def\blambda{\mbox{\boldmath $\lambda$}}
\def\bkappa{\mbox{\boldmath $\kappa$}}
\def\bmu{\mbox{\boldmath $\mu$}}
\def\bnu{\mbox{\boldmath $\nu$}}
\def\bsigma{\mbox{\boldmath $\sigma$}}
\def\bvarepsilon{\mbox{\boldmath $\varepsilon$}}
\def\bPi{\mbox{\boldmath $\Pi$}}
\def\bPhi{\mbox{\boldmath $\Phi$}}
\def\bGamma{\mbox{\boldmath $\Gamma$}}
\def\bLambda{\mbox{\boldmath $\Lambda$}}
\def\bOmega{\mbox{\boldmath $\Omega$}}
\def\bXi{\mbox{\boldmath $\Xi$}}
\def\bUpsilon{\mbox{\boldmath $\Upsilon$}}
\def\bcA{\mbox{\boldmath ${\mathcal A}$}}
\def\bcB{\mbox{\boldmath ${\cal B}$}}
\def\bcS{\mbox{\boldmath ${\cal S}$}}
\def\bcU{\mbox{\boldmath ${\cal U}$}}
\def\bcE{\mbox{\boldmath ${\cal E}$}}
\def\bcE{\pmb{\mathcal E}}
\def\bcF{\pmb{\mathcal F}}
\def\bcG{\pmb{\mathcal G}}
\def\bcT{\pmb{\mathcal T}}
\def\bcB{\pmb{\mathcal B}}
\def\bcD{\pmb{\mathcal D}}
\def\bcJ{\pmb{\mathcal J}}
\def\bcY{\pmb{\mathcal Y}}
\def\bcQ{\pmb{\mathcal Q}}
\def\bcH{\pmb{\mathcal H}}
\def\bcK{\pmb{\mathcal K}}
\def\bcV{\pmb{\mathcal V}}
\def\cF{{\mathcal F}}
\def\cG{{\mathcal G}}
\def\cH{{\mathcal H}}
\def\cL{{\mathcal L}}
\def\baq{\mbox{\bf [}}
\def\bcq{\mbox{\bf ]}}
\def\bu{{\bf u}}
\def\bv{{\bf v}}
\def\bx{{\bf x}}
\def\bp{{\bf p}}
\def\bb{{\bf b}}
\def\bc{{\bf c}}
\def\bg{{\bf g}}
\def\bj{{\bf j}}
\def\bog{{\bf g}}
\def\bX{{\bf X}}
\def\bE{{\bf E}}
\def\bd{{\bf d}}
\def\be{{\bf e}}
\def\bi{{\bf i}}
\def\bl{{\bf l}}
\def\bQ{{\bf Q}}
\def\bY{{\bf Y}}
\def\bF{{\bf F}}
\def\bT{{\bf T}}
\def\bG{{\bf G}}
\def\bD{{\bf D}}
\def\bB{{\bf B}}
\def\bH{{\bf H}}
\def\bI{{\bf I}}
\def\bK{{\bf K}}
\def\bL{{\bf L}}
\def\bO{{\bf O}}
\def\bP{{\bf P}}
\def\bM{{\bf M}}
\def\bm{{\bf m}}
\def\bq{{\bf q}}
\def\bJ{{\bf J}}
\def\bC{{\bf C}}
\def\bR{{\bf R}}
\def\bS{{\bf S}}
\def\bW{{\bf W}}
\def\bw{{\bf w}}
\def\bZ{{\bf Z}}
\def\bz{{\bf z}}
\def\bn{{\bf n}}
\def\bN{{\bf N}}
\def\bs{{\bf s}}
\def\ba{{\bf a}}
\def\bt{{\bf t}}
\def\by{{\bf y}}
\def\br{{\bf r}}
\def\bh{{\bf h}}
\def\bk{{\bf k}}
\def\bo{{\bf o}}
\def\sq{{\scriptscriptstyle q}}
\def\sA{{\scriptscriptstyle A}}
\def\sB{{\scriptscriptstyle B}}
\def\sC{{\scriptscriptstyle C}}
\def\sD{{\scriptscriptstyle D}}
\def\sE{{\scriptscriptstyle E}}
\def\sF{{\scriptscriptstyle F}}
\def\sK{{\scriptscriptstyle K}}
\def\sL{{\scriptscriptstyle L}}
\def\sM{{\scriptscriptstyle M}}
\def\sN{{\scriptscriptstyle N}}
\def\sT{{\scriptscriptstyle T}}
\def\sG{{\scriptscriptstyle G}}
\def\sO{{\scriptscriptstyle O}}
\def\sP{{\scriptscriptstyle P}}
\def\sQ{{\scriptscriptstyle Q}}
\def\sR{{\scriptscriptstyle R}}
\def\sS{{\scriptscriptstyle S}}
\def\sbT{{\scriptscriptstyle \bT}}
\def\bsN{\mbox{\boldmath ${\mathsf N}$}}
\def\bsM{\mbox{\boldmath ${\mathsf M}$}}
\def\bsK{\mbox{\boldmath ${\mathsf K}$}}
\def\bsH{\mbox{\boldmath ${\mathsf H}$}}
\def\bsX{\mbox{\boldmath ${\mathsf X}$}}
\def\bsF{\mbox{\boldmath ${\mathsf F}$}}
\def\bsG{\mbox{\boldmath ${\mathsf G}$}}
\def\bsC{\mbox{\boldmath ${\mathsf C}$}}
\def\bsA{\mbox{\boldmath ${\mathsf A}$}}
\def\bsB{\mbox{\boldmath ${\mathsf B}$}}
\def\bsP{\mbox{\boldmath ${\mathsf P}$}}
\def\bsI{\mbox{\boldmath ${\mathsf I}$}}
\def\bsL{\mbox{\boldmath ${\mathsf L}$}}
\def\bsJ{\mbox{\boldmath ${\mathsf J}$}}
\def\bs0{\mbox{\boldmath ${\mathsf O}$}}
\def\sbcE{{\pmb{\scriptscriptstyle \mathcal E}}} 
\def\bU{{\bf U}}
\def\bV{{\bf V}}
\def\bof{{\bf f}}
\def\vp{\varphi}
\def\bSig{{\bf Sigma}}
\def\bPsi{{\bf Psi}\!\!\!\!\!@!@!{\bf Psi}}
\def\cvd{$\qquad\Box$}
\def\vcvd{\hfill $\sqcap \!\!\!\!\sqcup$}
\def\bone{{\bf 1}}
\def\bzero{{\bf 0}}
\def\curl{\nabla \crux}
\def\dive{\nabla \cdot}
\def\circo{\buildrel\circ\over}
\def\diamo{\buildrel\diamond\over}
\def\bA{{\bf A}}
\def\nonimplica{\Longrightarrow \!\!\!\!\!\!\!\! /\quad}
\def\para{{\scriptscriptstyle{\parallel}}}
\def\perpe{{\scriptscriptstyle{\perp}}}
\def\Res{{\rm Res}}
\def\sinc{{\rm sinc}}
\def\scrE{\mathscr{E}}
\def\scrT{\mathscr{T}}
\def\scrS{\mathscr{S}}
\def\rR{{\rm R}}
\def\nablaR{\nabla\!_{\sR}\,}
\def\mT{{\mathfrak{T}}}
\def\mE{{\mathfrak{E}}}
\def\mS{{\mathfrak{S}}}
\def\msfS{\mbox{${\mathsf S}$}}
\def\strianup{{\scriptscriptstyle{\bigtriangleup}}}
\def\striandown{{\scriptscriptstyle{\bigtriangledown}}}
\def\triangoloup{\buildrel\strianup\over}
\def\triangolodown{\buildrel\striandown\over}
\def\SSQ#1#2{ {\vbox{\hrule height#2pt\hbox{\vrule width#2pt height#1pt
    \kern#1pt \vrule width#2pt }\hrule height#2pt}}\smallskip\relax\rm}
\def\cvd{\SSQ{2.5}{.4}}
\def\quadro{\buildrel\cvd\over}
\def\quadrato{\buildrel\cvd\over}
\def\triangolidown{\buildrel{\striandown\striandown}\over}
\def\sBox{{\scriptscriptstyle{\Box}}}
\def\quadri{\buildrel{\cvd\,\cvd}\over}
\def\tcr{\textcolor{red}}

\def\Sym{{\rm Sym}}
\def\sym{{\rm sym}}
\def\R{\mathbb R}
\def\N{\mathbb N}
\def\cR{{\mathcal R}}
\def\bxi{\mbox{\boldmath $\xi$}}
\def\bchi{\mbox{\boldmath $\chi$}}
\def\bkappa{\mbox{\boldmath $\kappa$}}
\def\bmu{\mbox{\boldmath $\mu$}}
\def\bnu{\mbox{\boldmath $\nu$}}
\def\bzeta{\mbox{\boldmath $\zeta$}}
\def\bGamma{\mbox{\boldmath $\Gamma$}}
\def\bb{{\bf b}}
\def\bh{{\bf h}}
\def\bx{{\bf x}}
\def\bv{{\bf v}}
\def\bq{{\bf q}}
\def\bA{{\bf A}}
\def\bX{{\bf X}}
\def\bD{{\bf D}}
\def\bE{{\bf E}}
\def\bF{{\bf F}}
\def\bL{{\bf L}}
\def\bQ{{\bf Q}}
\def\bT{{\bf T}}

\def\bone{{\bf 1}}
\def\bzero{{\bf 0}}
\def\tr{{\rm tr}\,}
\def\sE{{\scriptscriptstyle E}}
\def\sH{{\scriptscriptstyle H}}
\def\sK{{\scriptscriptstyle K}}
\def\sR{{\scriptscriptstyle R}}
\def\sT{{\scriptscriptstyle T}}
\def\sX{{\scriptscriptstyle X}}
\def\nablaR{\nabla\!_{\sR}\,}

\newtheorem{remark}{Remark}[section]
\newtheorem{proposition}{Proposition}[section]
\newtheorem{theorem}[proposition]{Theorem}
\newtheorem{corollary}[proposition]{Corollary}
\newtheorem{lemma}[proposition]{Lemma}

\date{}
\title{Modeling of heat conduction through rate
equations}

\author{Claudio Giorgi\thanks{DICATAM,  Universit\`a degli Studi di Brescia, Brescia, Italy. claudio.giorgi@unibs.it}, Angelo Morro\thanks{DIBRIS,  Universit\`a di Genova, via All'Opera Pia 13, Genova, 16145, Italy. angelo.morro@unige.it}, Federico Zullo\thanks{DICATAM,  Universit\`a degli Studi di Brescia, Brescia, Italy \& INFN, Sezione di Milano-Bicocca, Milano, Italy. federico.zullo@unibs.it}}


\maketitle


%


\begin{abstract}
 Starting from a classical thermodynamic approach, we derive rate-type equations to describe the behavior of heat flow in deformable media. Constitutive equations are defined in the material (Lagrangian)  description where the standard time derivative satisfies the principle of objectivity. 
The statement of the Second Law is formulated in the classical form and the thermodynamic restrictions are then developed following the Coleman-Noll procedure. However, instead of the Clausius Duhem inequality we consider the corresponding equality where the entropy production rate is prescribed by a non-negative constitutive function. Both the free energy and the entropy production are assumed to depend on a common set of independent variables involving, in addition to temperature, both temperature gradient and heat-flux vector together with their time derivatives.
This approach results in rate-type constitutive equations for the heat-flux vector that are intrinsically consistent with the Second Law and easily amenable to analysis. In addition to obtaining already known models (e.g., Cattaneo-Maxwell's, Jeffreys-like and Green-Naghdi's heat conductors), this scheme allows us to build new and more complex models of heat transport that may have applications in describing the thermal behavior in nanosystems. Indeed, when higher order time derivatives of the heat flux vector are involved, many different relaxation times occur within the rate equation.
 \end{abstract}

\section{Introduction}\label{sec1}



A huge class of heat conduction models beyond Fourier have been recently developed to account for relaxational and nonlocal effects, fast phenomena or wave propagation, such as being typical for biological systems, nanomaterials or nanosystems. Non-Fourier models mainly differ for their various thermodynamic backgrounds (thermodynamics of irreversible processes, extended irreversible thermodynamics, etc., see for instance \cite{CSJ_2010bis,JCVL,Lebon}). In connection with  wave propagation properties many models of heat propagation are presented  and discussed in \cite{Straughan}. Properties concerning their possible practical applications in light of experiments are debated in a recent review \cite{Kovacs}.
Despite the various approaches and procedures developed in the literature (see, e.g., \cite{SPKFVG,CSJ_2010,FGM_frac,Ch} and references tharein), the topic deserves further attention.
A challenging question is their possible compatibility with the Second Law. This article aims to discuss their deduction in the context of classical Continuum Thermodynamics and possible compatibility with the Second Law stated therein.
The question is of applicative interest, as constitutive equations must not only be consistent with the experimental data but must  also be in agreement with the fundamental laws of thermodynamics.

According to \cite{GZ_2023}, we develop a new approach to heat conduction theories that is inherently thermodynamic, as it originates directly from the Clausius-Duhem inequality. 
The  specific production of entropy $\sigma$ enters as a non-negative constitutive function, so that the Second Law is automatically satisfied.
The set of independent variables includes only the macroscopically observable fields and their temporal and spatial derivatives, without making any recourse to internal variables or ambiguous state variables (such as thermal displacement).  In particular, no constitutive prescription on the heat (or energy) influx $\bq$ is made, rather it is treated as an independent variable. 
Although other approaches treat both the heat flux and the temperature gradient as independent variables  (see, e.g., \cite{MCM, MorroJE} and references tharein) our scheme has several advantages: a material description is adopted  in order to avoid the problem of the objectivity of time derivatives, the rigidity assumption is not necessary and consistency with thermodynamics is much easier to prove than in theories where heat conduction involves histories \cite{Meccanica}, summed histories \cite{GP} or internal variables \cite{Van2015,Maugin}.

 The Coleman-Noll procedure \cite{CN63} is applied to derive thermodynamic restrictions on  the Helmholtz free energy $\psi$. 
 However, instead of the Clausius Duhem inequality we consider the corresponding equality where the entropy production rate $\sigma$  is prescribed by a non-negative constitutive function. Both the free energy and the entropy production are assumed to depend on a common set of independent variables.
 This approach is compatible with both rigid and deformable bodies.

In addition to obtaining already known models (e.g., Cattaneo-Maxwell's, Green-Naghdi's and Quintanilla's heat conductors),  this strategy, initially proposed in \cite{GZ_2023}, allows us to build new and more complex non-Fourier models of heat transport that have applications in describing the thermal behavior of materials where many different relaxation times occur.
  In particular in Sect.4 a class of rate-type linear constitutive equations of the second order for the heat flux is discussed in detail. The thermodynamic consistency of the model (originally presented in \cite{CR} and here referred to as Linear Second Order model) is proved here for the first time. In addition,  some features of the corresponding temperature equation are highlighted.


\section{Balance laws and the thermodynamic principles}\label{sec2}

 We consider a body occupying a three-dimensional time-dependent region $\Omega$.
We let $\cR$ be a chosen reference configuration, $\bX$ the vector position in $\cR$ of a material point and $\bchi(\bX, t)$ its motion with $t \in \R$ the time. Formally, $\Omega = \bchi(\cR, t)$. 
The velocity $\bv$ is given by the time derivative $\partial_t\bchi(\bX, t)$.
A superposed dot denotes the material time derivative, $\nabla$ is the gradient operator and then, for any $f(\bx, t)$, we have $\dot{f} = \partial_t f + \bv \cdot \nabla f$. Instead, $\nablaR$ is the gradient in the reference configuration. 

Let $\bF$ be the deformation gradient, $\bF (\bX, t)=\nablaR  \bchi(\bX, t)$ (in suffix notation $F_{i\sK} = \partial_{X_\sK} \chi_i$), satisfying the constraint $J:=\det \bF>0$, while 
 $\bL:=\nabla\bv$ is the velocity gradient, $L_{ij} = \partial_{x_j} v_i$, which is related to $\dot\bF$ as follows   \beq \bL=\dot\bF\bF^{-1}.
 \label{vel_grad} 
 \eeq
Also, $\bD = \sym \bL$ is the stretching where  $\sym$ denotes the symmetric part of a tensor and $\tr$ denotes the trace, so that $\tr\bD=\nabla\cdot\bv$.

Further, $\Sym$ denotes the set of symmetric second-order tensors and $\bone$ is the identity (or unit) tensor. For any $\bA\in\Sym$, $\bA > \bzero$ or $\bA \ge \bzero$ indicate that $\bA$ is positive definite or positive semi-definite, while $\Sym^+$ denotes the set of symmetric positive-definite tensors.

Let $\varepsilon$ be the internal energy density (per unit mass), $\bT$  the Cauchy stress, $\bq$ the heat flux vector, $\rho$ the mass density, $r$ the (external) heat supply and $\bb$ the mechanical body force per unit mass. 
The conservation of mass is expressed in local form by the continuity equation
\beq \dot{\rho} + \rho \nabla \cdot \bv = 0. \label{eq:rho}\eeq
The local form of the linear momentum and internal energy balance equations can be written as
\beq
\rho \dot{\bv} = \nabla \cdot \bT + \rho \bb ,
\label{eq:1motion}\eeq
\beq 
\rho \dot{\varepsilon} =\bT \cdot \bL - \nabla \cdot \bq + \rho r.\label{eq:en} \eeq

 Let  $\eta$ be the specific entropy function, ${\bh}$ the entropy-flux vector and $s$ the specific supply of entropy.
 All processes which are compatible with equations \eqref{eq:rho}-\eqref{eq:en} must satisfy the following entropy balance equation,
\[ \rho\, \dot\eta +\nabla\cdot {\bh} =  \rho s.\]
The classical local form, usually named after Clausius-Duhem, is obtained by letting
$$
{\bh}= \frac{\bq}{\theta},\qquad s=\frac{r}{\theta}+\gamma,
$$
where $\theta$ denotes the (positive) absolute temperature and the quantity $\gamma$ is referred to as {\it specific entropy production} \cite{CN63,Muller71,Muller85};
\begin{equation}
\label{loc_eta_ineq}
\rho \dot{\eta}  + \nabla\cdot\left( \frac{\bq}{\theta}  \right)- \frac{\rho r}{\theta}=\rho\gamma.
\end{equation}

In continuum thermodynamics the local form of the Second Law  is established by assuming 
\beq
\gamma(\bx,t)\ge0.
\label{strong_ineq}\eeq
 along any process compatible with all balance equations.
Furthermore, henceforth we assume that the entropy production $\gamma$ is given as a constitutive function of the common set of physical variables, as are the internal energy and entropy.  Conceptually our contribution follows the same scheme proposed in \cite{GN}, but imposing the fundamental prescription \eqref{strong_ineq} on all admissible processes (see also \cite[\S\,2.6]{GM_book}). 

Upon substitution of $\nabla\cdot \bq - \rho r$ from the energy equation (\ref{eq:en}) into \eqref{loc_eta_ineq} and multiplication by $\theta$ we obtain the basic thermodynamic relation 
\beq - \rho(\dot{\psi} + \eta \dot{\theta})  + \bT \cdot \bL 
- \frac 1 \theta \bq \cdot \nabla\theta =\rho\theta\gamma,\label{eq:CD0} \eeq
where $\psi= \varepsilon - \theta \eta$ denotes the specific {\it Helmholtz free energy}. 
Due to \eqref{strong_ineq}, equation \eqref{eq:CD0} becomes an inequality that must be satisfied along whatever process (see also \cite[eqn.(4.5)]{CN63}).
Finally, multiplying \eqref{eq:CD0} by $J$ and using the identity $\rho_\sR=J\rho$,
 we obtain the basic thermodynamic inequality in the material description
 \beq\begin{split}
-{\rho_\sR}\big(\dot{\psi} + \eta \dot{\theta}
\big)
+  \bT_{\sR\sR} \cdot \dot{\bE} 
- \frac 1 {\theta} \bq_\sR \cdot \nablaR \theta
=\rho_{\sR}\theta\gamma\ge0,
\end{split}\label{basic_ent_ineq}\eeq
where $\nablaR:=\bF^T\nabla$ and
\[ \bT_{\sR\sR} := J \bF^{-1} \bT \bF^{-T}, \qquad\bq_\sR := J \bF^{-1} \bq.
\]

\section{A new approach to rate-type constitutive equations}\label{sec3}

The revised Coleman-Noll procedure for the exploitation of the entropy principle leads to some consequences of the thermodynamic relation \eqref{basic_ent_ineq} after specifying three elements;
\begin{itemize}
\item[-] the set $\Xi_\sR$ of admissible variables,
\item[-] the free energy density function $\psi=\psi(\Xi_\sR)$,
\item[-] the entropy production function $\gamma=\gamma(\Xi_\sR)$.
\end{itemize}
The strategy can be illustrated first  in the case of elastic materials with heat conduction and viscosity. First we consider the {\it set of admissible variables}, $\Xi_\sR=(\theta, \bE, \nablaR \theta, \dot{\bE})$ and assume that both $\psi$ and $\gamma$ are functions of these variables. Upon evaluation of $\dot{\psi}$ and substitution in (\ref{basic_ent_ineq}) we obtain
\[ \begin{split} \rho_{\sR}(\partial_\theta \psi + \eta) \dot{\theta}
+ (\rho_{\sR} \partial_\bE \psi - \bT_{\sR\sR})\cdot \dot{\bE} 
+ \rho_{\sR} \partial_{\nablaR \theta} \psi \cdot \nablaR \dot{\theta}  &\\
+ \rho_{\sR} \partial_{\dot{\bE}}\psi \cdot \ddot{\bE}
 +\frac 1 \theta \bq_{\sR} \cdot \nablaR \theta &=-\rho_\sR\theta\gamma.\end{split} \] 
Neither $\psi$ nor $\gamma$ depend on  $\dot{\theta}$, $\ddot{\bE}$, therefore the linearity and arbitrariness of these variables imply
\beq \psi = \psi(\theta, \bE), \qquad  \eta=-\partial_\theta \psi. \label{eq:psi_0}\eeq
As a consequence, $\Sigma_\sR=(\theta, \bE)$ can be viewed as the {\it set of state variables}.
If $\gamma$ is independent of $\dot\bE$, that is, if the material is not viscous, then the linearity and arbitrariness of $\dot\bE$ also implies
\beq \bT_{\sR\sR}=\rho_{\sR}\partial_\bE \psi. \label{eq:stress_0}\eeq
Otherwise, the Clausius-Duhem relation for materials with heat conduction and viscosity reads
\[   (\rho_{\sR} \partial_\bE \psi - \bT_{\sR\sR})\cdot \dot{\bE} 
 +\frac 1 \theta \bq_{\sR} \cdot \nablaR \theta=-\rho_\sR\theta\gamma.\] 
Now, we have to specify the functions $\psi(\theta, \bE)$ and $\gamma(\theta, \bE, \nablaR \theta, \dot{\bE})$ to satisfy this equation and the fundamental requirement $\gamma\ge0$. For any choice of $\psi$, the simple quadratic function
\[\gamma= \bGamma_1(\theta, \bE)\dot\bE\cdot\dot\bE+\bGamma_2(\theta, \bE)\nablaR \theta\cdot\nablaR \theta,\]
where $\bGamma_1, \bGamma_2$ are positive semi-definite tensor-valued functions, satisfies $\gamma\ge0$ and implies
\[   (\rho_{\sR} \partial_\bE \psi + \bGamma_1\dot\bE- \bT_{\sR\sR})\cdot \dot{\bE} 
 +\Big(\frac 1 \theta \bq_{\sR}+ \bGamma_2\nablaR \theta\Big)\cdot \nablaR \theta=0.\] 
This equation holds identically if
\[
\bT_{\sR\sR}=\rho_{\sR} \partial_\bE \psi + \bGamma_1\dot\bE, \qquad  \bq_{\sR}=-\theta\bGamma_2\nablaR \theta,
\]
that represent constitutive equations for a Kelvin-Voigt viscoelastic material with Fourier heat conduction.
The choice of $\psi(\theta,\bE)$ uniquely determines the elastic component of the stress.

To describe this new approach in the case of materials of the rate type, we expand the basic set of Euclidean invariant variables by adding some quantities that are usually 
considered as constitutive functions; specifically, $\bT_{\sR\sR}$, $\bq_\sR$.  Hence we let
\beq \Xi_\sR := (\theta, \bE, \bT_{\sR\sR}, \bq_\sR, \nablaR \theta, \dot{\bE})\label{set_var}\eeq
be the set of admissible variables and assume that $\psi$, $\eta$, $\gamma$ are scalar-valued functions of $\Xi_\sR$.

 In view of the introduction of rate-type constitutive equations, we look for a scheme where $\dot \bq_\sR$ and $\nablaR \theta$, as well as $ \dot\bT_{\sR\sR}$ and $\dot{\bE}$, are regarded as mutually dependent variables. This actually implies that $\bT_{\sR\sR}$ and $\bE$ are implicitly dependent, just as happens in anholonomic system described by a set of  parameters subject to differential constraints that make their rates mutually dependent. 

Following this scheme, a wide range of rate-type models have been derived in recent papers \cite{MorroGiorgi_JELAS,MorroGiorgi_MDPI,GM_Materials}. In particular, nonlinear models for thermo-viscoelastic, viscoplastic and elastic-plastic materials (solids and fluids) subject to large deformations are established.


\subsection{Rate-type models of heat conduction}\label{sec3.1}

Lately ref.\,\cite{GZ_2023} developed a similar scheme to obtain several models of heat conduction in deformable solids.
The key idea in \cite{GZ_2023}  is to exploit  the formal similarity between scalar products
$$\bT_{\sR\sR} \cdot \dot{\bE},\qquad -\bq_\sR \cdot\nablaR(\ln \theta),$$
that appear in \eqref{basic_ent_ineq} and represent the mechanical and thermal powers of internal forces, respectively. 
By mimicking the procedure adopted in \cite{MorroGiorgi_MDPI, GM_Materials} with reference to mechanical power,  the exchange
\[ \bT_{\sR\sR}\ \longleftrightarrow\ \bq_\sR,\qquad \dot\bE\ \longleftrightarrow\ -\nablaR(\ln \theta),\] allows the construction of heat conduction models of the rate type, both known and new.

For definiteness, upon evaluation of $\dot{\psi}$ by virtue of \eqref{set_var} and substitution in (\ref{basic_ent_ineq}) we obtain
\[ \begin{split} \rho_{\sR}(\partial_\theta \psi + \eta) \dot{\theta}
+ (\rho_{\sR} \partial_\bE \psi - \bT_{\sR\sR})\cdot \dot{\bE} + \rho_{\sR} \partial_{\bT_{\sR\sR}} \psi \cdot \dot{\bT}_{\sR\sR} + \rho_{\sR} \partial_{ \bq_\sR}\psi \cdot { \dot\bq_\sR}& \\
+ \rho_{\sR} \partial_{\nablaR \theta} \psi \cdot \nablaR \dot{\theta}  
+ \rho_{\sR} \partial_{\dot{\bE}}\psi \cdot \ddot{\bE}
 +\frac 1 \theta \bq_{\sR} \cdot \nablaR \theta&=-\rho_\sR\theta\gamma.\end{split} \] 
The linearity and arbitrariness of $\dot{\theta}$, $\ddot{\bE}$, 
$\nablaR \dot{\theta}$, imply that 
\beq \psi = \psi(\theta, \bE, \bT_{\sR\sR},\bq_\sR), \qquad  \eta=-\partial_\theta \psi. \label{eq:psi_first_ord}\eeq
The functional dependence of $\psi$ suggests to define the set of {\it state variables} as
$$\Sigma_\sR=(\theta, \bE, \bT_{\sR\sR},\bq_\sR).$$
Accordingly, the thermodynamic inequality reduces to 
\beq  (\rho_{\sR} \partial_\bE \psi - \bT_{\sR\sR})\cdot \dot{\bE} + \rho_{\sR} \partial_{\bT_{\sR\sR}} \psi \cdot \dot{\bT}_{\sR\sR} + \rho_{\sR}\partial_{ \bq_\sR}\psi \cdot { \dot\bq_\sR}
+ \frac {\bq_{\sR}} \theta \cdot \nablaR \theta=-\rho_\sR\theta\gamma \le 0, \label{eq:ine0}\eeq 
where $\dot{\bT}_{\sR\sR}$, $  \dot{\bE}$ and $ \dot\bq_\sR$, $ \nablaR \theta$, respectively, are assumed to be implicitly dependent.
Since $\psi$ is independent of $\dot{\bq}_\sR, \nablaR \theta$,  then letting $\dot\bq_\sR=\nablaR \theta=\bzero$ we can  write \eqref{eq:ine0} in the form
\beq  \rho_{\sR} \partial_{\bT_{\sR\sR}} \psi \cdot \dot{\bT}_{\sR\sR}+(\rho_{\sR} \partial_\bE \psi - \bT_{\sR\sR})\cdot \dot{\bE} =-\rho_\sR\theta\gamma^{\sE\sT} \le 0, 
 \label{eq:ineETq}\eeq
where $\gamma^{\sE\sT}$ is the entropy production density $\gamma$ when $\dot\bq_\sR=\nablaR\theta = \bzero$. 
Likewise,  
\beq 
\rho_{\sR}\partial_{ \bq_\sR}\psi \cdot { \dot\bq_\sR}
+  {\bq_{\sR}} \cdot \nablaR(\ln \theta)=-\rho_\sR\theta\gamma^{q}\le 0,
\label{eq:ineETqq}\eeq 
where $\gamma^{q}$ is the entropy production density  when $\dot{\bT}_{\sR\sR}=\dot{\bE} = \bzero$. 
Furthermore, we let 
$$\gamma = \gamma^{\sE\sT} + \gamma^{q}.$$  
The entropy productions $\gamma^{\sE\sT}$ and $\gamma^{q}$, as well as $\gamma$, are nonnegative constitutive functions to be determined according to the constitutive model.

Disregarding heat conduction and exploiting inequality (\ref{eq:ineETq}), memory properties of viscoelasticity, elastoplasticity and viscoplasticity were modeled with suitable nonlinear rate-type stress-strain relations in \cite{MorroGiorgi_JELAS,MorroGiorgi_MDPI}. 
On the contrary, to establish rate-type models of heat conduction we limit our attention to (\ref{eq:ineETqq}) and neglect the dependence of constitutive functions  on ${\bE}$ and ${\bT}_{\sR\sR}$. 

A given rate-type model of heat conduction involving a set of variables $\Xi_\sR$  is said to be {\it consistent with thermodynamics} if there exists at least a pair of functions $\psi(\Xi_\sR), \gamma^q(\Xi_\sR)$ that satisfy the inequality (\ref{eq:ineETqq}).
For instance, the Maxwell-Cattaneo-Vernotte  (MCV) model (see \cite{Cattaneo58,Vernotte})
\beq
\tau{\dot\bq}_\sR+{\bq}_\sR=-\bkappa\nablaR \theta,\qquad \bkappa\in\Sym ,
\label{eq:MCV}\eeq
is proved to be consistent with thermodynamics by letting (see \cite[\S\,4.2]{GZ_2023})
\[
\rho_{\sR}\psi=\rho_{\sR}\psi_0(\theta)+\frac\tau{2\theta}\bq_\sR\cdot\bkappa^{-1}\bq_\sR, \qquad \rho_\sR\gamma^q =\frac1{\theta^2}\bq_\sR\cdot\bkappa^{-1}\bq_\sR,
\]
where $\bkappa$ must be positive-definite in order to have $\gamma^q\ge0$.
The sign of $\tau$ is not prescribed by thermodynamic arguments. However, the common assumption $\tau>0$ implies that $\psi$ has a minimum at ${\bq}_\sR = \bzero$; $\tau$ is called {\it relaxation time} and the Fourier law is recovered as $\tau\to 0^+$.


\subsection{First order rate-type models}\label{sec3.2}

Following the scheme devised in \cite{GZ_2023}, hereafter we neglect all variables involving stress and strain, but we expand the previously considered set of admissible variables by adding first-order time derivatives of $ \bq_\sR$ and $ \nablaR \theta$.
Hence we let 
\[ \Xi_\sR := (\theta, \bq_\sR, \dot{\bq}_\sR, \nablaR \theta, \nablaR \dot\theta).\]
Moreover, let $\psi, \eta, \gamma$ be dependent on $\Xi_\sR$. Upon evaluation of $\dot{\psi}$ and substitution in (\ref{basic_ent_ineq}) we obtain\footnote{Hereafter, in agreement with neglecting the dependence of constitutive functions  on ${\bE}$ and ${\bT}_{\sR\sR}$, we disregard the mechanical power $\bT_{\sR\sR}\cdot\dot\bE$. Consequently, $\gamma$ stands for $\gamma^q$, the entropy production density occurring when $\dot{\bT}_{\sR\sR}=\dot{\bE} = \bzero$.}
\[ \begin{split} \rho_{\sR}(\partial_\theta \psi + \eta) \dot{\theta}  
+ \rho_{\sR}\partial_{ \bq_\sR}\psi \cdot { \dot\bq_\sR}+\rho_{\sR}\partial_{ \dot\bq_\sR}\psi \cdot { \ddot\bq_\sR}& \\
  + \rho_{\sR} \partial_{\nablaR \theta} \psi \cdot \nablaR \dot{\theta}+ \rho_{\sR} \partial_{\nablaR \dot\theta} \psi \cdot \nablaR \ddot{\theta}+\frac 1 \theta \bq_{\sR} \cdot \nablaR \theta&=-\rho_\sR\theta\gamma.\end{split} \] 
The linearity and arbitrariness of $\dot \theta$ and $\nablaR \ddot{\theta}$  imply 
\[ \psi = \psi(\theta,\bq_\sR, \dot\bq_\sR, \nablaR \theta), \qquad \eta=-\partial_\theta \psi , \]
so that the set of state variables turns out to be
$$\Sigma_\sR:=(\theta,\bq_\sR, \dot\bq_\sR, \nablaR \theta)$$
 and the entropy inequality reduces to 
\beq \rho_{\sR}\partial_{ \bq_\sR}\psi \cdot { \dot\bq_\sR}+\rho_{\sR}\partial_{ \dot\bq_\sR}\psi \cdot { \ddot\bq_\sR}  + \rho_{\sR} \partial_{\nablaR \theta} \psi \cdot \nablaR \dot{\theta}
+ \frac {\bq_{\sR}} \theta \cdot \nablaR \theta=-\rho_\sR\theta\gamma \le 0. \label{eq:ine1}\eeq 
In view of the rate-type models considered below, for any value assigned to the state variables, the derivatives $\ddot \bq_\sR$ and  $\nablaR \dot\theta$ must be regarded as mutually dependent. This in turn implies that $\dot \bq_\sR$ and $\nablaR \theta$ are implicitly dependent.

Exploiting this procedure  in \cite{GZ_2023} the thermodynamic consistency of some rate-type heat conduction models was demonstrated. For completeness, the results obtained in this article are summarized below.
A first set of results involving well-known models was derived by neglecting the dependence of $\psi$ on $ \dot \bq_\sR$.


\subsubsection{Green-Naghdi type III (GN III) heat conductors}

Derivation with respect to time of the linear constitutive equation of a type III heat conductor according to Green and Naghdi \cite{GN3} yields 
\beq\dot\bq_\sR=-\bxi\nablaR \theta- \bkappa\nablaR \dot\theta,\qquad  \bxi,\bkappa\in\Sym.
\label{eq:GNIII}\eeq
where $\bxi,\bkappa$ are constant tensors. A differential version of the Fourier law is obtained when $\bxi=\bzero$.
If we let (see \cite[\S\,5.2]{GZ_2023})
\[ \rho_\sR\psi = \rho_\sR\psi_0(\theta) + 
\frac{1}{2\theta } [\bq_{\sR}+ \bkappa\nablaR \theta]\cdot\bxi^{-1} [\bq_{\sR}+ \bkappa\nablaR \theta],\qquad \rho_\sR \gamma =\frac1{\theta^2}\nablaR \theta\cdot \bkappa\nablaR \theta,\]
thermodynamic consistency is ensured by $\bkappa\in \Sym^+$. However $\bxi$ is required to be invertible in order to guarantee the boundedness of $\psi$.


\subsubsection{Heat conductors of the Jeffreys type}
The constitutive equation of a heat conductor of the Jeffreys type is given by
\beq
\tau\dot \bq_\sR+\bq_\sR=-\bkappa\nablaR \theta- \tau\bzeta\nablaR \dot\theta,\qquad  \bkappa,\bzeta\in\Sym.
\label{eq:Jeffreys}\eeq
The Fourier law is recovered as $\tau\to 0^+$. 
The Jeffreys type conductor can be obtained as a combination of two different models. Let $\bq_\sR^{(1)}$, $\bq_\sR^{(2)}$ be heat fluxes governed by the Fourier law and the MCV law, respectively,
$$\bq_\sR^{(1)}=- \bkappa_1\nablaR \theta  \qquad \tau\bq_\sR^{(2)}+\bq_\sR^{(2)}=-\bkappa_2\nablaR \theta  ,$$
where $\bkappa_1,\bkappa_2$ are positive-semidefinite second-order tensors. Hence
$\dot\bq_\sR^{(1)}=- \bkappa_1\nablaR \dot\theta $. 
It follows that
\[
\tau\big(\dot \bq_\sR^{(1)}+\dot \bq_\sR^{(2)}\big)+\bq_\sR^{(1)}+ \bq_\sR^{(2)}=-(\bkappa_1+\bkappa_2)\nablaR \theta- \tau\bkappa_1\nablaR \dot\theta,
\]
Consequently, the flux $\bq_\sR=\bq_\sR^{(1)}+ \bq_\sR^{(2)}$ satisfies \eqref{eq:Jeffreys} with $\bkappa=\bkappa_1+\bkappa_2$ and $\bzeta=\bkappa_1$. 

As proved in \cite[\S\,5.1]{GZ_2023}, different choices of $\psi$ and $\gamma$ allow the model to be consistent with thermodynamics. For instance, either
\[\begin{split} \rho_\sR\psi_1 &= \rho_\sR\psi_0(\theta) + 
\frac{\tau}{2\theta } [\bq_{\sR}+ \bzeta\nablaR \theta]\cdot(\bkappa+\bzeta)^{-1} [\bq_{\sR}+ \bzeta\nablaR \theta],\\
\rho_\sR \gamma_1 &=\frac1{\theta^2}\bq_\sR\cdot(\bkappa+\bzeta)^{-1} \bq_\sR+\frac1{\theta^2}\nablaR \theta\cdot \bzeta( \bkappa+\bzeta)^{-1}\bkappa\nablaR \theta.\end{split}\]
or
\[\begin{split}  \rho_\sR\psi_2&= \rho_\sR\psi_0(\theta) + 
\frac{\tau}{2\theta } [\bq_{\sR}+ \bzeta\nablaR \theta]\cdot(\bkappa-\bzeta)^{-1} [\bq_{\sR}+ \bzeta\nablaR \theta],\\
\rho_\sR \gamma_2& =\frac1{\theta^2}(\bq_\sR +\bzeta\nablaR \theta\big)\cdot(\bkappa-\bzeta)^{-1} (\bq_\sR +\bzeta\nablaR \theta\big)+\frac1{\theta^2}\nablaR \theta\cdot \bzeta\nablaR \theta.\end{split}\]
In the former case, the thermodynamic consistency, $\gamma_1\ge0$, is ensured if and only if $\bkappa\in\Sym^+$ and $\bzeta=\beta\bkappa$, $\beta\ge0$. In the latter, $\gamma_2\ge0$ is ensured if and only if $\bkappa>\bzeta\ge0$.

\medskip
When the dependence of $\psi$ on $ \dot \bq_\sR$ is allowed, new models of heat conductors consistent with thermodynamics were obtained in \cite{GZ_2023}.


\subsubsection{Quintanilla's heat conduction model}
A new theory of thermoelasticity phenomena have been proposed by Quintanilla in \cite{Quintanilla} by modifying the Green-Naghdi's type III theory.
Deriving with respect to time the original equation, we obtain the rate-type Quintanilla model
\beq
\tau\ddot \bq_\sR+\dot \bq_\sR=-\bxi\nablaR \theta- \bkappa\nablaR \dot\theta,\qquad  \bxi,\bkappa\in\Sym.\label{eq:Q}
\eeq
It can be verified (see \cite[\S\,5.3]{GZ_2023}) that the free energy  and entropy production related to the Quintanilla model respectively  take the form
\[ \begin{split}
\rho_\sR\psi& = \rho_\sR\psi_0(\theta) +\frac{1}{2\theta}\big[\tau\dot\bq_{\sR}+\bkappa\nablaR \theta\big]\cdot{\bkappa}{(\bkappa-\tau\bxi)^{-1}} \bxi^{-1} \big[\tau\dot\bq_{\sR}+\bkappa\nablaR \theta\big]
\\&+\frac{1}{2\theta}\Big( \bq_{\sR}
+ 2[\tau\dot{\bq}_{\sR}+\bkappa \nablaR \theta ]\Big)\cdot\bxi^{-1}  \bq_{\sR}, \\
\rho_\sR \gamma&=\frac{1}{\theta^2}(\tau\dot\bq_{\sR}+\bkappa\nablaR \theta)\cdot(\bkappa-\tau\bxi)^{-1} (\tau\dot\bq_{\sR}+\bkappa\nablaR \theta).
\end{split}\]
Assuming $\tau>0$, thermodynamic consistency is ensured by $\bkappa>\tau\bxi$. However $\bxi$ is required to be invertible in order to guarantee the boundedness of $\psi$.
Letting $\tau\to0$ equation \eqref{eq:Q}  reduces to \eqref{eq:GNIII}, and the free energy  and  entropy production of the GN  III conductor are recovered. Hence, the Quintanilla model represents a proper generalization of the GN III linear theory.

In order to study the wave propagation, we restrict our attention to a unidimensional rigid body.
We then couple equation \eqref{eq:Q}  with the internal energy balance law and obtain
\[
\rho_\sR c_v\dot \theta+\partial_{\sX}q_\sR=\rho_\sR r, \qquad \tau\ddot q_\sR+\dot q_\sR=-\xi\partial_{\sX} \theta- \kappa\partial_{\sX}\dot \theta.\]
Letting $r=0$, we obtain a linear version of the well-known Moore-Gibson-Thompson equation \cite{Quintanilla},
 \[
\tau \dddot\theta + \ddot\theta=\frac{1}{\rho_\sR c_v}\left( \xi\partial_{\sX} ^2 \theta+\kappa\partial_{\sX} ^2 \dot\theta\right).
\]

\subsubsection{Heat conductors of the Burgers type}
To our knowledge, Burgers-type heat conductors were first proposed in \cite{GZ_2023}. They are characterized by the rate-type equation
\beq
\lambda\ddot \bq_\sR+\tau\dot \bq_\sR+\bq_\sR=-\bkappa\nablaR \theta- \tau\bzeta\nablaR \dot\theta.\label{Burg}
\eeq
By analogy with the rheological model of the Burgers fluid, this equation can be obtained by considering a mixture of two components, each  characterized by a conduction mechanism described by the MCV equation \eqref{eq:MCV}.
The Burgers-like model  with $\lambda,\tau>0$ is thermodynamically consistent if and only if one of the following hypotheses occurs (see \cite[\S\,5.4]{GZ_2023})
\begin{itemize}
  \item[i)]  $\bkappa=0$, $\bzeta\in\Sym^+$;
   \item[ii)] $\bkappa\in\Sym^+$, $\tau^2\bzeta\ge\lambda\bkappa$.
\end{itemize}
Due to the linearity of the model equation \eqref{Burg}, both $\psi$ and $\gamma$ are quadratic functions of the state variables $\bq_\sR, \dot\bq_\sR, \nablaR \theta$ (see \cite[eqns.~(47) and (60)]{GZ_2023}). These functions are not unique.

As to the propagation of thermal waves, we consider a rigid unidimensional body.  If the specific heat supply $r$ vanishes, the resulting system
\[
\rho_\sR c_v\dot \theta+\partial_{\sX}q_\sR=0, \qquad \lambda\ddot q_\sR+\tau\dot q_\sR+q_\sR=-\kappa\partial_{\sX} \theta- \tau\zeta\partial_{\sX}\dot \theta.\]
 leads to the Joseph-Preziosi temperature equation \cite{JP},
\beq
\lambda \dddot\theta + \tau \ddot\theta+\dot\theta=\frac{1}{\rho_\sR c_v}\left( \kappa\partial_{\sX}^2 \theta+\tau\zeta\partial_{\sX}^2 \dot\theta\right).
\label{JP_temp_eq}\eeq


\section{Higher order  rate-type models of heat conduction}\label{sec4}

To describe some new rate-type models in heat conduction we consider  higher-order time derivatives of the heat flux vector and the temperature gradient in the set of admissible variables, namely
\[ \Xi_\sR := (\theta, \bq_\sR, \dot{\bq}_\sR,  \ddot{\bq}_\sR,  \nablaR \theta, \nablaR \dot\theta, \nablaR \ddot\theta).\]
Letting $\psi, \eta,\gamma$ be dependent on $\Xi_\sR$, upon evaluation of $\dot{\psi}$ and substitution in (\ref{basic_ent_ineq}), we obtain
\[ \begin{split} \rho_{\sR}(\partial_\theta \psi + \eta) \dot{\theta}
  + \rho_{\sR}\partial_{ \bq_\sR}\psi \cdot { \dot\bq_\sR}+\rho_{\sR}\partial_{ \dot\bq_\sR}\psi \cdot { \ddot\bq_\sR} +\,\rho_{\sR}\partial_{ \ddot\bq_\sR}\psi \cdot { \dddot\bq_\sR}& \\
  + \rho_{\sR} \partial_{\nablaR \theta} \psi \cdot \nablaR \dot{\theta}+ \rho_{\sR} \partial_{\nablaR \dot\theta} \psi \cdot \nablaR \ddot{\theta}+\rho_{\sR} \partial_{\nablaR \ddot\theta} \psi \cdot \nablaR \dddot{\theta}+\frac 1 \theta \bq_{\sR} \cdot \nablaR \theta&=-\rho_\sR\theta\gamma.\end{split} \] 
Hence, the linearity and arbitrariness of $\dot{\theta}$, 
$\nablaR \dddot{\theta}$  imply that 
\[\psi = \psi(\theta,\bq_\sR, \dot\bq_\sR,\ddot\bq_\sR, \nablaR \theta,\nablaR \dot{\theta}), \qquad \eta=-\partial_\theta \psi ,\]
and the entropy inequality reduces to 
\beq \begin{split}
 \rho_{\sR}\partial_{ \bq_\sR}\psi \cdot { \dot\bq_\sR}+\rho_{\sR}\partial_{ \dot\bq_\sR}\psi \cdot { \ddot\bq_\sR} +\,\rho_{\sR}\partial_{ \ddot\bq_\sR}\psi \cdot { \dddot\bq_\sR}&\\
  + \rho_{\sR} \partial_{\nablaR \theta} \psi \cdot \nablaR \dot{\theta}
 + \rho_{\sR} \partial_{\nablaR \dot\theta} \psi \cdot \nablaR \ddot{\theta}
+ \frac {\bq_{\sR}} \theta \cdot \nablaR \theta&=-\rho_\sR\theta\gamma \le 0.
 \end{split} \label{eq:ine2}\eeq 

Let $\Sigma_\sR=(\theta,\bq_\sR, \dot\bq_\sR,\ddot\bq_\sR, \nablaR \theta,\nablaR \dot{\theta})$ be the set of state variables. As previously remarked, for any value assigned to the state variables, the derivatives $\dddot \bq_\sR$ and  $\nablaR \ddot\theta$ must be regarded as mutually dependent. This in turn implies that $\ddot \bq_\sR$ and $\nablaR \dot\theta$, as well as $\dot \bq_\sR$ and $\nablaR \theta$, are implicitly dependent. However, constitutive models in which the free energy is independent of some variables of $\Sigma_\sR$  can also be considered.

In particular, we are interested here to some special models where the dependence of $\psi$ on $ \ddot \bq_\sR$ is neglected. If this is the case, inequality (\ref{eq:ine2}) becomes
\beq \begin{split}
 \rho_{\sR}\partial_{ \bq_\sR}\psi \cdot { \dot\bq_\sR}+\rho_{\sR}\partial_{ \dot\bq_\sR}\psi \cdot { \ddot\bq_\sR} 
  + \rho_{\sR} \partial_{\nablaR \theta} \psi \cdot \nablaR \dot{\theta}&\\
 + \rho_{\sR} \partial_{\nablaR \dot\theta} \psi \cdot \nablaR \ddot{\theta}
+ \frac {\bq_{\sR}} \theta \cdot \nablaR \theta&=-\rho_\sR\theta\gamma \le 0.
 \end{split} \label{eq:ine3}\eeq 


\subsection{A linear second-order model (LSO)}\label{sec4.1}

A model of heat conductor is considered in the form (see  \cite{CR})
\beq
\lambda\ddot \bq_\sR+\tau\dot \bq_\sR+\bq_\sR=-\bmu\nablaR \theta- \tau\bnu\nablaR \dot\theta- \lambda\bkappa\nablaR \ddot\theta.\label{eq:CR}
\eeq
This model represents an extension of the Burgers-type conductor to which it reduces when $ \bkappa=\bzero$. On the other hand, when $\lambda=0$ it reduces to the Jeffreys model. 

The LSO model can be obtained,  by considering a mixture of three different substances and assuming that the resulting heat flux vector is given by the sum $\bq_\sR=\bq_\sR^{(1)}+\bq_\sR^{(2)}+\bq_\sR^{(3)}$.
In the first component the heat conduction follows the Fourier law, whereas the second and third components are
characterized by a conduction mechanism described by the Maxwell-Cattaneo equation (\ref{eq:MCV}), namely
\[
\begin{split}
\bq_\sR^{(1)}=-\bkappa^{(1)}\nablaR \theta,
\qquad
 \tau_2\dot\bq_\sR^{(2)}+ \bq_\sR^{(2)}=-\bkappa^{(2)}\nablaR \theta,
 \qquad
  \tau_3\dot\bq_\sR^{(3)}+ \bq_\sR^{(3)}=-\bkappa^{(3)}\nablaR \theta.
\end{split}
\]
We recover \eqref{eq:CR} after some manipulations by letting
\[
\tau=\tau_2+\tau_3,\ \ \lambda=\tau_2\tau_3, \\ \  \bmu=\bkappa^{(1)}+\bkappa^{(2)}+\bkappa^{(3)}, \ \  \bnu=\bkappa^{(1)}+\frac1\tau\Big(\tau_3\bkappa^{(2)}+\tau_2\bkappa^{(3)}\Big),\ \  \bkappa=\bkappa^{(1)}.
\]

In order to investigate the consistency of the LSO constitutive equation
(\ref{eq:CR}) with inequality (\ref{eq:ine3}), we restrict our attention to the isotropic case where $\bmu=\mu\bone$, $\bnu=\nu\bone$ and $\bkappa=\kappa\bone$.
Moreover, we assume 
$$ \lambda,\kappa\neq0$$
 to exclude the Jeffreys and Burgers conductors. Accordingly we can consider $\ddot{\bq}_{\sR}$ as a linear function of   ${\bq}_{\sR}, \dot{\bq}_{\sR}, \nablaR \theta , \nablaR \dot\theta$ and $ \nablaR \ddot\theta$, namely
\beq
\ddot{\bq}_{\sR}=-\frac\tau\lambda\dot \bq_\sR-\frac1\lambda\bq_\sR-\frac\mu\lambda\nablaR \theta- \frac{\tau\nu}\lambda\nablaR \dot\theta- \kappa\nablaR \ddot\theta.
\label{eq:BSE_iso}\eeq
 Upon substitution for $\ddot{\bq}_{\sR}$ from (\ref{eq:BSE_iso}) into \eqref{eq:ine3}, we have
\[\begin{split} 
\rho_{\sR}\Big(\partial_{ \bq_\sR}\psi -\frac\tau\lambda\partial_{ \dot\bq_\sR}\psi \Big)\cdot { \dot\bq_\sR}
-\rho_{\sR}\frac1\lambda\partial_{ \dot\bq_\sR}\psi \cdot  \bq_\sR
+ \Big(\frac {\bq_{\sR}} \theta -\rho_{\sR} \frac\mu\lambda\partial_{ \dot\bq_\sR}\psi\Big) \cdot\nablaR \theta
& \\
+\rho_{\sR}\Big( \partial_{\nablaR \theta} \psi-\frac{\tau\nu}\lambda\partial_{ \dot\bq_\sR}\psi \Big)\cdot\nablaR \dot\theta 
+\rho_{\sR}\Big( \partial_{\nablaR \dot\theta} \psi-\kappa\partial_{ \dot\bq_\sR}\psi \Big)\cdot\nablaR \ddot\theta 
&=-\rho_\sR\theta\gamma \le 0.\end{split}\]
Since $\psi$ is independent of $\nablaR \ddot\theta$, assuming that $\gamma$ is also independent, the linearity and arbitrariness of $\nablaR \ddot\theta$ imply 
\beq \partial_{\nablaR \dot\theta} \psi=\kappa\partial_{ \dot\bq_\sR}\psi. \label{eq:resCR}\eeq
Otherwise, we can assume the constraint \eqref{eq:resCR} and in turn obtain that $\gamma$ is independent of $\nablaR \ddot\theta$.
Anyway, \eqref{eq:ine3} reduces to
 \beq\begin{split} 
 \rho_{\sR}\Big(\partial_{ \bq_\sR}\psi -\frac\tau\lambda\partial_{ \dot\bq_\sR}\psi \Big)\cdot { \dot\bq_\sR}
-\rho_{\sR}\frac1\lambda\partial_{ \dot\bq_\sR}\psi \cdot  \bq_\sR &\\
+ \Big(\frac {\bq_{\sR}} \theta -\frac\mu\lambda\rho_{\sR} \partial_{ \dot\bq_\sR}\psi\Big) \cdot\nablaR \theta
+\rho_{\sR}\Big( \partial_{\nablaR \theta} \psi-\frac{\tau\nu}\lambda\partial_{ \dot\bq_\sR}\psi \Big)\cdot\nablaR \dot\theta
&=-\rho_\sR\theta\gamma \le 0.
\end{split}\label{eq:redCR}  \eeq
As suggested by the linearity of the model, the free energy $\psi$ is assumed to have the following quadratic expression,
\begin{equation}\label{free_en}\begin{split}  \rho_\sR\psi &= \rho_\sR\psi_0(\theta) +\frac{\alpha_1}2 |\bq_{\sR}|^2 +\frac{\alpha_2}2 |\dot\bq_{\sR}|^2+\frac{\alpha_3}2 |\nablaR \theta|^2+\frac{\alpha_4}2 |\nablaR \dot\theta|^2+ \beta_1{\bq}_{\sR} \cdot \dot\bq_{\sR} \\
& +\beta_2\bq_{\sR} \cdot \nablaR \theta +\beta_3\dot\bq_{\sR} \cdot \nablaR \theta+\beta_4\bq_{\sR} \cdot \nablaR \dot\theta+\beta_5\dot\bq_{\sR} \cdot \nablaR \dot\theta+\beta_6\nablaR \theta \cdot \nablaR \dot\theta,\end{split}\end{equation}
whence
\[\begin{split}\rho_\sR\partial_{{\bq}_{\sR}}\psi &= \alpha_1 \bq_{\sR}+\beta_1 \dot\bq_{\sR}+\beta_2\nablaR \theta+\beta_4\nablaR\dot \theta, \\
\rho_\sR\partial_{\dot\bq_{\sR}}\psi &=\beta_1 \bq_{\sR}+\alpha_2 \dot\bq_{\sR}+\beta_3\nablaR \theta+\beta_5\nablaR\dot \theta,  \\
\rho_\sR\partial_{\nablaR \theta}\psi &=\beta_2 \bq_{\sR}+\beta_3 \dot\bq_{\sR}+\alpha_3\nablaR \theta+
\beta_6\nablaR\dot \theta, \\
\rho_\sR\partial_{\nablaR\dot \theta}\psi &=\beta_4 \bq_{\sR}+\beta_5 \dot\bq_{\sR}+\beta_6\nablaR \theta+\alpha_4\nablaR \dot\theta.
\end{split}\]
Upon substitution into \eqref{eq:resCR}  we obtain
\beq 
\beta_4 =\kappa\beta_1, \qquad\beta_5 =\kappa\alpha_2 , \qquad \beta_6=\kappa\beta_3,\qquad \alpha_4=\kappa\beta_5.
\label{eq:CR_1}\eeq
Likewise, from \eqref{eq:redCR} and \eqref{eq:CR_1} it follows
\beq\begin{split} 
A_{11} |\bq_{\sR}|^2+A_{22} |\dot\bq_{\sR}|^2+A_{33} |\nablaR \theta|^2+A_{44} |\nablaR \dot\theta|^2+2A_{12}\bq_{\sR}\cdot\dot\bq_{\sR}+2A_{13}\bq_{\sR}\cdot\nablaR \theta\\
+2A_{14}\bq_{\sR}\cdot\nablaR \dot\theta+2A_{23}\dot\bq_{\sR}\cdot\nablaR \theta+2A_{24}\dot\bq_{\sR}\cdot\nablaR \dot\theta+2A_{34}\nablaR\theta\cdot\nablaR \dot\theta=\rho_\sR\theta \gamma \ge 0, 
\end{split} \label{eq:CR_2}\eeq
where
\small
\[\begin{split} 
 &  A_{11}= \frac{\beta_1}{\lambda},\;   A_{22}= \frac{\tau\alpha_2-\lambda\beta_1}{\lambda},\; A_{33}= \frac{\mu\beta_3 }{\lambda},\; 
 A_{44}=\frac{\tau\nu\kappa\alpha_2}{\lambda} -\kappa\beta_3  \\
&  A_{12}= \frac{\alpha_2+\tau\beta_1-\lambda\alpha_1}{2\lambda},\; A_{13}=\frac{\beta_1\mu\theta+\beta_3 \theta-\lambda}{2\lambda\theta},\; A_{14}= \frac{\beta_1\nu\tau+\kappa\alpha_2-\lambda\beta_2}{2\lambda}, \\  
&   A_{23}= \frac{\mu\alpha_2+\tau\beta_3 -\lambda\beta_2}{2\lambda}, \; A_{24}=\frac{\tau(\kappa+\nu)\alpha_2-\lambda(\kappa\beta_1+\beta_3 )}{2\lambda}, \; A_{34}=  \frac{\kappa\mu\alpha_2+\nu\tau\beta_3 -\lambda\alpha_3}{2\lambda}.  
\end{split}\]

\normalsize
So overall matrix $A$ is characterized by 6 unknowns, $\alpha_i,\beta_i$, $i=1,2,3$,   5 real material parameters, $\lambda,\tau,\mu,\nu,\kappa$ and  $\theta$. 
To ensure thermodynamic consistency, we look for the conditions on these material parameters so that the symmetric 4-by-4 matrix $  A$ is positive semidefinite,  
 i.e. all principal minors of $A$ are nonnegative (see, for instance \cite[\S\,7.6]{CDM}). Naturally, there may be different values of the unknowns that are compatible with these conditions. This is related to the fact that there can exist different free energy functions that are consistent with thermodynamics.

The discussion of the  thermodynamic consistency of the model is too cumbersome to be included in the body of the paper. We therefore postpone it to the Appendix.

\subsection{LSO temperature equation}\label{sec4.2}
Let us look briefly at properties of the solutions of the temperature equation corresponding to the LSO model. For simplicity, let us consider a  rigid body. Without an external energy supply, the energy balance gives 
\begin{equation}
\rho_\sR c_v\dot\theta = - \nablaR \cdot \bq_\sR.
\end{equation}
After combining this equation with (\ref{eq:CR}) and assuming the body to be isotropic, we get
\beq
\rho_\sR c_v\left(\lambda\dddot\theta+\tau\ddot\theta+\dot\theta\right)=\mu\nabla_\sR^2 \theta+ \tau\nu\nabla_\sR^2 \dot\theta+ \lambda\kappa\nabla_\sR^2 \ddot\theta.\label{eq:CR1}
\eeq
Letting $\theta(\bx, t)= T(t)Y(\bX)$, we obtain
\beq
\rho_\sR c_v\left(\lambda\dddot T+\tau\ddot T+\dot T\right)Y=\left(\mu T+ \tau\nu \dot T+ \lambda\kappa \ddot T\right)\nabla_\sR^2 Y\label{eq:CR2}
\eeq
Equation (\ref{eq:CR2}) can be separated: the spatial variable $Y$ solves the Helmholtz equation 
\begin{equation*}\label{eqH}
\nabla_\sR^2 Y=-\Lambda Y
\end{equation*}
where $\Lambda$ is a constant. Once the domain and boundary conditions have been fixed, the Helmholtz equation possesses non-trivial solutions only if $\Lambda$ assumes specific values (the eigenvalues). It is well-known that under the most common boundary conditions the differential operator  $-\nabla_\sR^2$ is strictly positive selfadjoint with discrete spectrum. Hence,  its eigenvalues are non-negative, countably infinite and not bounded by any constant value. So, let us denote their set  with $\{\Lambda_n\}_{n\in \N}$, ordered in an ascending sequence, i.e. $\Lambda_n<\Lambda_{n+1}$. Equation \eqref{eq:CR2} for $T(t)$ then reads
\begin{equation}\label{T}
\lambda \dddot T + (\tau+\tilde{\Lambda}_n\lambda\kappa)\ddot T+(1+\tilde{\Lambda}_n\tau\nu)\dot T+\tilde{\Lambda}_n\mu T=0,
\end{equation}
where we set $\tilde{\Lambda}_n:= {\Lambda_n}/{\rho_\sR c_v}$. If $T(t) \propto e^{wt}$, then $w$ is a root of the cubic equation
\begin{equation}\label{T1}
\lambda w^3 + (\tau+\tilde{\Lambda}_n\lambda\kappa)w^2+(1+\tilde{\Lambda}_n\tau\nu)w+\tilde{\Lambda}_n\mu =0.
\end{equation}

To avoid solutions diverging at infinity, we consider only  decaying or oscillating solutions to equation \eqref{T}. Therefore, we look for necessary and sufficient conditions under which all roots of \eqref{T1} have negative real parts.
According to the Routh-Hurwitz criterion, all the coefficients must have the same sign and the product of the coefficients of $w$ and $w^2$ minus the product of the coefficients of $w^3$ and $w^0$ must be positive.  For simplicity, we assume the positivity of $\lambda$ and $\tau$, a condition that could be deduced from the physical assumptions we made to build the model. The application of the Routh-Hurwitz criterion to (\ref{T1}) yields
\begin{equation}\label{reles}
\lambda,\tau >0, \;\tau+\tilde{\Lambda}_n\lambda\kappa >0, \; 1+\tilde\Lambda_n\tau\nu>0,\; \mu>0,\; (\tau +\tilde{\Lambda}_n\lambda\kappa)(1+\tilde\Lambda_n\tau\nu)-\tilde\Lambda_n\lambda\mu>0,
\end{equation}
 for any $n \in \N$. By exploiting the unboundedness of $\tilde{\Lambda}_n$, the second and third inequalities give $\kappa \ge0$ and $\nu \ge0$, respectively. So all material  parameters  must be non negative; in particular
 $$\lambda, \tau, \mu>0,\qquad  \kappa, \nu\ge0.$$
  The last inequality  is quadratic with respect to  $\tilde\Lambda_n$ and reads
\begin{equation}\label{relp}
\lambda\kappa\tau\nu\tilde\Lambda_n^2+(\tau^2\nu+\lambda(\kappa-\mu))\tilde\Lambda_n+\tau>0.
\end{equation}

If $ \kappa, \nu>0$, then the coefficient of $\tilde\Lambda_n^2$ is positive. Since all the values of $\tilde\Lambda_n$ are non negative, the previous relation is satisfied either if all the coefficients  of the corresponding quadratic equation are positive (in this case any real roots are negative), or if the discriminant of the corresponding quadratic equation is negative (in this case (\ref{relp}) is satisfied for any real value of $\tilde\Lambda_n$). In the first case $\tau^2\nu+\lambda(\kappa-\mu)>0$ is required, i.e.
\begin{equation}\label{mu1}
\mu<\kappa+\frac{\tau^2\nu}{\lambda}.
\end{equation}
In the second case we let $\left(\tau^2\nu+\lambda(\kappa-\mu)\right)^2-4\tau^2\nu\lambda\kappa <0$,
which implies
\begin{equation}\label{mu2}
\mu \in \frac{1}{\lambda}\left(\left(\sqrt{\lambda\kappa}-\sqrt{\tau^2\nu}\right)^2, \left(\sqrt{\lambda\kappa}+\sqrt{\tau^2\nu}\right)^2\right).
\end{equation}
From the combination of (\ref{mu1}) and (\ref{mu2}) follows that (\ref{relp}) is satisfied when
\begin{equation}\label{mupr}
\mu < \frac{\left(\sqrt{\lambda\kappa}+\sqrt{\tau^2\nu}\right)^2}{\lambda}.
\end{equation}

If $ \kappa=0, \nu>0$, then
\begin{equation}\label{relp1}
(\tau^2\nu-\lambda\mu)\tilde\Lambda_n+\tau>0
\end{equation}
and the unboundedness of $\tilde{\Lambda}_n$ yields $\tau^2\nu\ge\lambda\mu$. 

If $ \kappa>0, \nu=0$, then
\begin{equation}\label{relp2}
\lambda(\kappa-\mu)\tilde\Lambda_n+\tau>0.
\end{equation}
and the unboundedness of $\tilde{\Lambda}_n$ yields $\kappa\ge\mu$.

Let us consider now the limit case $\mu=0$.  If $\mu=0$, we see that one of the roots of (\ref{T1}) is zero, the other two being given by the equation
\begin{equation}\label{T3}
\lambda w^2 + (\tau+\tilde{\Lambda}_n\lambda\kappa)w+(1+\tilde{\Lambda}_n\tau\nu) = 0.
\end{equation}
The roots of (\ref{T3}) must be negative, meaning that all the coefficients in (\ref{T3}) must have the same sign. This implies
\begin{equation}\label{reps1}
\lambda >0, \;  (\tau+\tilde{\Lambda}_n\lambda\kappa) >0, \; (1+\tilde{\Lambda}_n\tau\nu)>0. 
\end{equation}
As we have seen, the conditions on the coefficient of $w$ and on the coefficient of $w^0$ imply $\kappa >0$ and $\nu >0$. So we have that all the parameters $\lambda$, $\tau$, $\kappa$, $\nu$ must be positive and $\mu=0$.

Finally, we notice that a special behavior occurs when \eqref{relp} holds as an equality for some $\bar n\in\N$, i.e.
\begin{equation}\label{relp3}
\lambda\kappa\tau\nu\tilde\Lambda_{\bar n}^2+(\tau^2\nu+\lambda(\kappa-\mu))\tilde\Lambda_{\bar n}+\tau=0,
\end{equation}
and inequalities (\ref{reles}) still hold. In this case, equation (\ref{T1}) has a negative real solution and a couple of purely imaginary solutions,
\begin{equation}
w_1=-\frac{\tilde\Lambda_{\bar n}\lambda\kappa+\tau}{\lambda}, \quad w_{\pm}=\pm \text{i}\omega=\pm \text{i}\sqrt{\frac{\tilde\Lambda_{\bar n}\tau\nu+1}{\lambda}}.
\end{equation}
To fix ideas, we consider a rigid body in an environment with a given constant temperature $\theta_0$. Then solutions of the Cauchy problem (\ref{eq:CR1}) can be written as:
\begin{equation}
\theta(\bx, t)= \theta_0 + \sum_{n=1}\sum_{i=1}^3c_n^i e^{w_n^i t}Y_n(\bX),
\end{equation}
where the constants $c_n^i$ are fixed by the initial conditions. Notice that the boundary conditions $\left.Y_n(\bX)\right|_{\partial\Omega}=0$, where $\partial\Omega$ are the boundaries of the body, fix the values of the eigenvalues $\Lambda_n$. In the limit $t \to \infty$ only the terms proportional to $e^{w_{\pm}t}$ survive and we get
\begin{equation}
\lim_{t\to \infty}\theta(\bx, t)= \theta_0 + \left(\alpha_{\bar n}\sin(\omega t)+\beta_{\bar n}\cos(\omega t)\right)Y_{\bar n}(\bX)
\end{equation}
for two suitable constants $\alpha_{\bar n}$ and $\beta_{\bar n}$. Notice that the condition (\ref{relp3}) is very peculiar, since it is equivalent to say that it exists a value of $\tilde\Lambda_{\bar n}$ such that $\mu$ can be expressed as
\begin{equation}
\mu = \frac{\left(\sqrt{\lambda\kappa}+\sqrt{\tau^2\nu}\right)^2}{\lambda}+\frac{\left(\tilde\Lambda_{\bar n}\sqrt{\kappa\lambda\nu\tau}-\sqrt{\tau}\right)^2}{\tilde\Lambda_{\bar n}\lambda}.
\end{equation}
\section{Conclusion}\label{sec5}
In these Conclusions we would like to make a comment about the simultaneous thermodynamic and dynamic consistency of the LSO model, i.e. we are asking under which conditions the LSO model has a nonnegative entropy production (see Appendix \ref{secA1}) and the evolution of the temperature equation is described by bounded functions eventually approaching the equilibrium steady state. Firstly, let us consider the case $\mu>0$. By comparing the items in table  (\ref{tab1}) with the results given in Sect.\ref{sec4.2}, we see that items 10 and 12-16 can be excluded, since in all these cases $\lambda <0$, contrary to the hypotheses (\ref{reles}). Also, items 4-7 must be excluded if, according to (\ref{reles}), $\mu >0$. Moreover, it is worth noting that items 9 and 11 are very specific: the possible values of the parameter $\mu$ are described by the intersection of three different intervals and this intersection must satisfy also (\ref{mupr}): if the intersection of the three intervals is empty and/or the condition (\ref{mupr}) is not satisfied the model would be inconsistent. Finally, we observe that items 1-3 and 8 are instead all consistent with (\ref{reles}) and (\ref{mupr}); hence we conclude that these choices of the material parameters are consistent both from a thermodynamic and dynamic point of view.

Particular attention should be paid to the case $\mu=0$. This is compatible only with items 4-7 in table (\ref{tab1}). However, items 6 and 7 must be excluded since $\lambda<0$, contrary to the hypotheses (\ref{reps1}). On the contrary, the remaining cases 4 and 5 agree with condition (\ref{reps1}) and are therefore consistent both from a thermodynamic and dynamic point of view.

%


%
%
%

\section*{Acknowledgements}

The research leading to this work has been developed under the auspices of Universit\`a degli Studi di Brescia, Universit\`a degli Studi di Genova and INDAM-GNFM. F.Z. acknowledges also the support of INFN, Gr. IV - Mathematical Methods in NonLinear Physics.


%
%
%
%
%
\section*{Appendices}

\section{Conditions for the thermodynamic consistency of the LSO model}\label{secA1}

\small
For further convenience, we adopt the following notations. The 3-by-3 principal submatrices of $A$ are obtained by deleting just the $h$-th row and column of the matrix. Their determinants, called 3-by-3 principal minors, are referred to as $d_{h}$, $h=1,2,3$.
Accordingly, the 2-by-2 principal submatrices of $A$ are obtained by deleting the $h$-th and $k$-th rows and columns and their determinants are denoted by $d_{h,k}$, $h\neq k$. The 1-by-1 principal submatrices of $A$  coincide with the elements of its principal diagonal.

We start by looking at the 2-by-2 principal minor
\begin{equation}\label{d24}
d_{1,3}:=\det
\begin{pmatrix}
A_{22}    &   A_{24} \\
  A_{24}    &  A_{44}
\end{pmatrix} = -\frac{\left[\lambda(\beta_3 -\kappa\beta_1)+(\kappa-\nu)\tau\alpha_2\right]^2}{4\lambda^2}
\end{equation}
 If the numerator does not vanish, $d_{1,3}$ takes a negative value; therefore, we are forced to impose $d_{1,3}=0$. We use this condition together with (\ref{d24}) to fix the value of $\beta_3 $, namely
\begin{equation}\label{g4}
\beta_3 =\kappa\beta_1+\tau\alpha_2\frac{\nu-\kappa}{\lambda}.
\end{equation}
Taking into account this value of $\beta_3 $, we consider the 3-by-3 principal minors $d_1$ and $d_3$ where the element $A_{22}$ can be factored out,
\begin{equation}\label{d3}
d_{3}:=\det
\begin{pmatrix}
A_{11}    &   A_{12} & A_{14} \\
  A_{12}    &  A_{22} & A_{24} \\
A_{14}  & A_{24}  & A_{44}
\end{pmatrix} = -A_{22}\frac{\left(\lambda\beta_2-\kappa\lambda\alpha_1+(\kappa-\nu)\tau\beta_1\right)^2}{4\lambda^2}
\end{equation}
\begin{equation}\label{d1}
d_{1}:=\det
\begin{pmatrix}
A_{22}    &   A_{23} & A_{24} \\
  A_{23}    &  A_{33} & A_{34} \\
A_{24}  & A_{34}  & A_{44}
\end{pmatrix} = -A_{22}\frac{\left(\lambda^2(\alpha_3-\kappa\beta_2)+\kappa\lambda\beta_1(\kappa-\nu)\tau-\alpha_2(\kappa-\nu)^2\tau^2\right)^2}{4\lambda^4}.
\end{equation}
Notice that $A_{22}$ is itself a 1-by-1 minor, so it must be non-negative. As a consequence, either 
$A_{22}=0$, or $A_{22}\neq0$ and the numerators of $d_3$ and $d_1$ must vanish.

\subsection*{$A_{22}=0$} Condition $A_{22}=0$  leads to 
\begin{equation}\label{g1}
\beta_1=\frac{\tau}{\lambda}\alpha_2.
\end{equation}
From (\ref{g4}) and (\ref{g1}) it follows
\[\beta_3 =\kappa\frac{\tau\alpha_2}{\lambda}+\tau\alpha_2\frac{\nu-\kappa}{\lambda}=\frac{\tau\nu}{\lambda}\alpha_2,
\]
and  $A_{24}=A_{44}=0$. So we consider the following 2-by-2 minors
\begin{equation}\label{di4}
d_{2,3}:=\det
\begin{pmatrix}
A_{11}    &   A_{14} \\
  A_{14}    &  A_{44}
\end{pmatrix} = -A_{14}^2, \quad 
d_{1,2}:=\det
\begin{pmatrix}
A_{33}    &   A_{34} \\
  A_{34}    &  A_{44}
\end{pmatrix} = -A_{34}^2.
\end{equation}
The elements $A_{14}$ and $A_{34}$ then must vanish, so yielding
\[
\beta_2=\frac{\nu\tau^2+\kappa\lambda}{\lambda^2}\alpha_2, \qquad \alpha_3=\frac{\lambda\kappa\mu+{\tau^2\nu^2}}{\lambda^2}\alpha_2 .
\]
 As a consequence of (\ref{g1}),   $A_{\textrm{i}4}=A_{4\textrm{i}}=0$, $\textrm{i}=1,\dots,4$, and the entropy production (\ref{eq:CR_2}) reduces to a 3-by-3 matrix as in  the Burger's model.
 Applying \eqref{g1} and \eqref{g4}  and deleting the fourth row and column of the matrix $A$, we obtain
\begin{equation}\label{eq_61special}\begin{split} 
 &  A_{11}= \frac{\tau}{\lambda^2}\alpha_2,\;   A_{22}= 0,\; A_{33}= \frac{\mu\tau\nu}{\lambda^2}\alpha_2, \\
&  A_{12}= \frac{({\lambda}+{\tau^2})\alpha_2-\lambda^2\alpha_1}{2\lambda^2},\; A_{13}=\frac{{\tau}\theta(\mu+\nu)\alpha_2-\lambda^2}{2\lambda^2\theta}, \;   A_{23}= \frac{\mu-\kappa}{2\lambda}\alpha_2.  
\end{split}\end{equation}
If $\tau=0$ then $d_{2,4}=-1/(2\theta)^2<0$ and $A$ cannot be positive semidefinite.  Accordingly we let $\tau\neq0$ and consider 
\[
d_{3,4}:=\det
\begin{pmatrix}
A_{11}    &   A_{12} \\
  A_{12}    &  A_{22}
\end{pmatrix} = -A_{12}^2
\]
which yields $A_{12}=0$ so that
\[
\alpha_1=\frac{{\lambda}+{\tau^2}}{\lambda^2}\alpha_2.
\]
Then we consider 
\[
d_{1,4}:=\det
\begin{pmatrix}
A_{22}    &   A_{23} \\
  A_{23}    &  A_{33}
\end{pmatrix} = -A_{23}^2=- \frac{(\mu-\kappa)^2}{4\lambda^2}\alpha^2_2.
\]
Hence, either $\mu-\kappa=0$ or $\alpha_2=0$.
Since
\[
d_{2,4}:=\det
\begin{pmatrix}
A_{11}    &   A_{13} \\
  A_{13}    &  A_{33}
\end{pmatrix} = A_{11}A_{33}-A_{13}^2=\frac{\mu\tau^2\nu}{\lambda^4}\alpha^2_2-
\left[\frac{{\tau}\theta(\mu+\nu)\alpha_2-\lambda^2}{2\lambda^2\theta}\right]^2
\]
we infer that $d_{2,4}<0$ when $\alpha_2=0$, so we are forced to assume $\alpha_2\neq0$ and 
$$\mu=\kappa.$$ This  gives
\[
d_{2,4}=\frac{\kappa\tau^2\nu}{\lambda^4}\alpha^2_2-
\left[\frac{{\tau}\theta(\kappa+\nu)\alpha_2-\lambda^2}{2\lambda^2\theta}\right]^2
\]
from which it follows that $d_{2,4}\ge0$ if and only if
\begin{equation}
0\ge
{\tau^2}\theta^2(\kappa-\nu)^2\alpha^2_2-2\lambda^2{\tau}\theta(\kappa+\nu)\alpha_2+\lambda^4.
\label{ineq_alpha_2}\end{equation}
which is a quadratic inequality involving the unknown $\alpha_2$. 

First we discuss the case $\nu=\mu=\kappa\neq0$ so that \eqref{ineq_alpha_2} reduces to 
$$\kappa\tau\alpha_2\ge\lambda^2/4\theta.$$
In this case, $A_{11}$ and $A_{33}$ are positive provided that $\kappa>0$.
Accordingly, thermodynamic consistency is achieved if $\mu=\nu=\kappa>0$ and $\tau\neq0$. 

%
%

Otherwise let $\nu\neq\kappa$ and 
$$\Delta=\lambda^4{\tau^2}\theta^2[(\kappa+\nu)^2-(\kappa-\nu)^2]=4\kappa\nu\lambda^4{\tau^2}\theta^2.$$
When $\Delta<0$ the inequality \eqref{ineq_alpha_2} is false for any value of $\alpha_2$. Accordingly, we assume $\kappa\nu\ge0$. Recalling that $\kappa\neq0$, we discuss two items: $\nu=0$ and $\kappa\nu>0$.
\begin{itemize}
  \item $\mu=\kappa\neq0$, $\tau\neq0$ and $\nu=0$: then $\Delta=0$ and \eqref{ineq_alpha_2} is satisfied provided that 
  $\alpha_2={\lambda^2}/{{\tau}\theta\kappa}$.
As a consequence $A_{13}=0$  and $A_{11}= {1}/{\theta\kappa}$
is the only non null term of the matrix. Accordingly, thermodynamic consistency is achieved if $\kappa>0$. 
  \item $\mu=\kappa\neq0$, $\tau\neq0$, $\nu\neq\kappa$ and $\kappa\nu>0$: then $\Delta>0$. In particular either $\kappa,\nu>0$ or $\kappa,\nu<0$.  
  \begin{itemize}
  \item If  $\kappa,\nu>0$ then \eqref{ineq_alpha_2} is satisfied provided that $\alpha_2\in(\alpha^-_2,\alpha^+_2)$ where
   \[
  \alpha^\pm_2=\lambda^2\frac{(\sqrt{\kappa}\pm\sqrt{\nu} )^2}{\tau\theta(\kappa-\nu)^2}.
  \]
  Hence, $\tau\alpha_2>0$ and then $A_{11}= \frac{\tau}{\lambda^2}\alpha_2>0$, $A_{33}= \frac{\kappa\nu}{\lambda^2}\tau\alpha_2>0$. In addition $A_{12}=A_{22}=A_{23}=0$ and $d_{1,4}=d_{3,4}=d_4=0$, $d_{2,4}\ge0$ so that $A$ turns out to be positive semidefinite and  thermodynamic consistency is achieved.
  \item If  $\kappa,\nu<0$ \eqref{ineq_alpha_2} is satisfied provided that  
$\alpha_2\in(\alpha^-_2,\alpha^+_2)$ where
   \[
  \alpha^\pm_2=-\lambda^2\frac{(\sqrt{|\kappa|}\pm\sqrt{|\nu|} )^2}{\tau\theta(\kappa-\nu)^2}.
  \]
 Hence, $\tau\alpha_2<0$ and then thermodynamic consistency cannot be achieved.
\end{itemize}
 \end{itemize}
Summarizing this item, the LSO model is thermodynamically consistent if 
$$\boxed{\mu=\kappa>0, \qquad\tau\neq0, \qquad\nu\ge0}.$$

\subsection*{$A_{2,2}\neq 0$} 
In the following the combination $\nu-\kappa$ will appear frequently, so for ease in writing we define 
$$x=\nu-\kappa$$
and  we replace the parameter $\nu$ with $\nu=x+\kappa$.
Since $\lambda\neq 0$, equalities $d_1=0$ and $d_3=0$ can be used to determine the values of $\alpha_3$ and $\beta_2$,  respectively. So we set
$$
\alpha_3=\frac{x^2\tau^2\alpha_2+2x\kappa\lambda\beta_1\tau+\kappa^2\lambda^2\alpha_1}{\lambda^2}, \quad \beta_2=\kappa\alpha_1+\frac{x\tau\beta_1}{\lambda}.
$$
Taking into account these values, we now consider the following 2-by-2 minor
\begin{equation}\label{d13}\begin{split}
& d_{2,4}:=\det
\begin{pmatrix}
A_{11}    &   A_{13} \\
  A_{13}    &  A_{33}
\end{pmatrix} =\\
&= -\left(\frac{\beta_1}{2\lambda}\right)^2\mu^2+\frac{\beta_1\mu}{2\lambda^3\theta}\left(\theta(\kappa\lambda\beta_1+x\tau\alpha_2)+\lambda^2\right)-\frac{\left(\theta(\kappa\lambda\beta_1+x\tau\alpha_2)-\lambda^2\right)^2}{4\lambda^4\theta^2}
\end{split}\end{equation}
Let $\Delta_{2,4}$ denotes the discriminant of $d_{2,4}$ with respect to $\mu$. We find
\begin{equation}\label{delta13}
\Delta_{2,4}=\frac{\beta_1^2}{\lambda^4\theta}\left(\lambda\kappa\beta_1+x\tau\alpha_2\right)
\end{equation}
Looking at  (\ref{d13}), if $\mu\beta_1=0$  we are forced to set $d_{2,4}=0$. Otherwise, if $\beta_1\mu\neq 0$ the sign of the quadratic polynomial in $\mu$ depends on $\Delta_{2,4}$. So we consider two main subcases: i) $\mu\beta_1=0$ and ii) $\mu\beta_1 \neq 0$. 
\begin{itemize}
 \item[\bf i) ] $\mu\beta_1=0$. This case has three subitems: i1) $\mu\neq 0$ and $\beta_1 =0$,  i2) $\mu = 0$ and $\beta_1 =0$ and i3) $\mu= 0$ and $\beta_1 \neq0$. 
 \begin{itemize}
\item[\bf i.1)] $\mu\neq 0$ and $\beta_1=0$. In this case we are forced to set
$
\lambda^2-x\tau\alpha_2\theta=0,
$
otherwise $d_{2,4}$ is negative. 
Accordingly, $x\tau$ and $\alpha_2$ must be different from zero since $\lambda\neq 0$. 
From $\beta_1=0$ it follows $A_{11}=0$ which in turn implies $d_{3,4}=-A_{12}^2$ and  $d_{2,3}=-A_{14}^2$. So we are forced to set $A_{12}=A_{14}=0$. These conditions give the values of $\alpha_1$ and $\alpha_2$ in terms of the other parameters and  temperature as
\[
\alpha_{1}=\frac{\lambda}{x\tau\theta}, \quad \alpha_2=\frac{\lambda^2}{x\tau\theta}.
\]
At this point all the elements $A_{1i}$, $i=1,...,4$, and all the 3-by-3 minors are equal to zero. The diagonal elements $A_{ii}$, $i=1,2,3$, are 
non-negative provided that 
$$x\lambda>0,\qquad\mu > 0.$$
 The 2-by-2 minors different from zero are  $d_{1,2}$ and
$d_{1,4}=\kappa^2d_{1,2}$, so we require $d_{1,4}\geq 0$, namely
\begin{equation}\label{d14n}
d_{1,4}=-\frac{1}{4x^2\tau^2\theta^2}\left(\lambda^2\mu^2-2\lambda(\lambda\kappa+x\tau^2)\mu+(\lambda\kappa-x\tau^2)^2\right)\geq 0.
\end{equation}
The numerator of $d_{1,4}$  is a second order  polynomial in $\mu$ whose  discriminant  is  given by $16\kappa x\tau^2\lambda^3$. In order to satisfy (\ref{d14n}), this discriminant must be non-negative, a condition which implies $\kappa\geq 0$. The interval of admissible values is given by $\mu \in [\mu_1^{-}, \mu_1^{+}]$, with
$$
\mu_1^{\pm}=\left(\sqrt{\kappa}\pm |\tau|\sqrt{\frac{x}{\lambda}}\right)^2.
$$

\item[\bf i.2)] $\mu = 0$ and $\beta_1=0$. In this case both $A_{11}$ and $A_{33}$ are equal to zero. Then 
\[
d_{3,4}=-\frac{(\alpha_2-\lambda\alpha_1)^2}{4\lambda^2}, \;\; d_{2,4}=-\frac{(x\tau\alpha_2\theta-\lambda^2)^2}{4\lambda^4\theta^2}, \;\; d_{1,4}=-\frac{(x\tau^2\alpha_2-\kappa\lambda^2\alpha_1)^2}{4\lambda^4}.
\]
As before, we assume $x\tau$ and $\alpha_2$ different from zero otherwise $d_{2,4}<0$. Moreover  we must have $d_{3,4}=0$, $d_{2,4}=0$, $d_{1,4}=0$ from which it follows
\[
\kappa=\frac{x\tau^2}{\lambda},\;\; \alpha_1=\frac{\lambda}{x\tau\theta},\;\;  \alpha_2=\frac{\lambda^2}{x\tau\theta}.
\]
Recalling that $x=\nu-\kappa$, the first equality yields
$$\nu=\frac{\kappa(\lambda+\tau^2)}{\tau^2}, \qquad x=\frac{\kappa\lambda}{\tau^2}.$$
By inserting these values in the matrix $A$, only four elements do not vanish, namely
\[
A_{22}=\frac{\tau^2}{\kappa\theta}, \;\;A_{24}=A_{42}=\frac{\tau^2}{\theta}, \;\; A_{44}=\frac{\kappa\tau^2}{\theta},
\]
and the matrix $A$ is positive semi-definite iff $\kappa>0$.

\smallskip
\item[\bf i.3)] $\mu = 0$ and $\beta_1\neq0$. In this case the element $A_{33}$ vanishes. Since $d_{2,4}=-A_{13}^2$, $d_{1,4}=-A_{23}^2$, $d_{1,2}=-A_{34}^2=-\kappa^2 A_{23}^2$, we must set $A_{13}=0$ and $A_{23}=0$. These two equations can be written explicitly as
\[
(\kappa\lambda\beta_1+x\tau\alpha_2)\theta-\lambda^2=0, \;\; x\tau^2\alpha_2+\beta_1(\kappa-x)\tau\lambda-\kappa\alpha_1\lambda^2=0.
\]
We use these relations to fix the values of $\alpha_1$ and $\beta_1$ in terms of the other variables,
\[
\alpha_1=\frac{\tau\left(x^2\tau\alpha_2\theta+\lambda^2(\kappa-x)\right)}{\kappa^2\lambda^2\theta},\;\;\; \beta_1=\frac{\lambda^2-x\tau\alpha_2\theta}{\kappa\lambda\theta}.
\]
The matrix elements different from zero are given by:
\[\begin{split}
& A_{11}=\frac{\lambda^2-x\tau\alpha_2\theta}{\kappa\lambda^2\theta},\;\; A_{12}=\frac{\alpha_2}{2\lambda}-\frac{x\tau}{2\kappa\lambda}A_{22},\;\; A_{14}=\kappa A_{12},\\
& A_{22}=\frac{\tau(\kappa+x)\alpha_2\theta-\lambda^2}{\kappa\lambda\theta}, \;\; A_{24}=\kappa A_{22}, \;\;A_{44}=\kappa^2 A_{22}.
\end{split}\]
The  1-by-1 minors must be non negative. The remaining minors different from zero are $d_{3,4}$ and $d_{2,3}=\kappa^2d_{3,4}$; $d_{3,4}$ is given by
\[\begin{split}
d_{3,4}=&-\frac{\left(x\tau^2 (\kappa+x)+\lambda\kappa^2\right)^2}{4\lambda^4\kappa^4}\alpha_2^2\\&
+\frac{\tau\left(x^2\tau^2(\kappa+x)
+\lambda\kappa^2(2\kappa+3x)\right)}{2\lambda^2\kappa^4\theta}\alpha_2-\frac{x^2\tau^2+4\lambda\kappa^2}{4\kappa^4\theta^2}
\end{split}\]
{\bf a)} First we analyze the possibility $x\tau^2 (\kappa+x)+\lambda\kappa^2\neq 0$.  In this case the discriminant of $d_{3,4}$ with respect to $\alpha_2$ must be non negative in order to have a non negative value for $d_{3,4}$. So we get
\[
\Delta_{3,4}=\frac{\tau^2(\kappa+x)-\lambda\kappa}{\kappa^3\lambda^2\theta}\geq 0.
\]
After replacing $x=\nu-\kappa$ this condition is equivalent to
\[
 \lambda\leq \frac{\tau^2\nu}\kappa.
\]
 Admissible values of $\alpha_2$ are fixed by inequality $d_{3,4}\geq 0$, i.e. $\alpha_2\in I_{\alpha_2}=[\alpha_2^{-}, \alpha_2^+]$,
\[
\alpha_2^{\pm}=\frac{\lambda^2\left(x^2\tau^3(\kappa+x)+\lambda\kappa^2\tau(2\kappa+3x)\pm2\kappa^2\sqrt{\kappa\lambda^2(\tau^2(\kappa+x)-\lambda\kappa)}\right)}{\theta\left(x\tau^2 (\kappa+x)+\lambda\kappa^2\right)^2}
\]
If $\alpha_2$ belongs to the interval $I_{\alpha_2}$ then  $A_{11}$ and $A_{22}$ are non negative provided $\kappa>0$. Indeed we have
\[\begin{split}
A_{11}&=\frac{\lambda^2-x\tau\alpha_2^{\pm}\theta}{\lambda^2\kappa\theta}= \frac{\left(\lambda\kappa^2\mp x\tau\sqrt{\kappa\left(\tau^2(\kappa+x)-\lambda\kappa\right)}\right)^2}{\left(x\tau^2 (\kappa+x)+\lambda\kappa^2\right)^2\kappa\theta},\\
A_{22}&=\frac{\tau(\kappa+x)\alpha_2^{\pm}\theta-\lambda^2}{\lambda\kappa\theta}=\frac{\kappa\lambda^2\left((x+k)\tau\pm\sqrt{\kappa\left(\tau^2(\kappa+x)-\lambda\kappa\right)}\right)^2}{\left(x\tau^2 (\kappa+x)+\lambda\kappa^2\right)^2\theta}.
\end{split}\]
{\bf b)} The analysis of the sub-case ii.3) is completed by considering the possibility
\[
\lambda=-\frac{x(\kappa+x)\tau^2}{\kappa^2}.
\]
Notice that $x(\kappa+x)\tau \neq 0$ since $\lambda\neq 0$. Now, the relation $d_{3,4}\ge0$ becomes
\[
-\frac{\alpha_2}{x\tau\theta}+\frac{x\tau^2(4\kappa+3x)}{4\kappa^4\theta^2}\geq 0.
\]
-- If $x\tau>0$ then we have
\begin{equation}\label{a2l}
\alpha_2\leq \frac{x^2\tau^3(4\kappa+3x)}{4\kappa^4\theta}.
\end{equation}
The corresponding values of $A_{11}$ and $A_{22}$ are positive only if $\kappa>0$. Indeed, according to (\ref{a2l}), let us assume 
$$
\alpha_2=\frac{x^2\tau^3(4\kappa+3x)}{4\kappa^4\theta}-N,
$$ 
where $N$ is a suitable non-negative quantity. Then we have
\[
A_{11} = \frac{(x+2\kappa)^2}{4\theta\kappa(x+\kappa)^2}+\frac{\tau x N}{\kappa \lambda^2},\;\; A_{22}=\frac{\kappa N}{\tau x} + \frac{x^2\tau^2}{4\kappa^3\theta},
\]
showing that $A_{11}$ and $A_{22}$ are non-negative iff $\kappa >0$.

-- If $x\tau<0$ one has
\[
\alpha_2\geq \frac{x^2\tau^3(4\kappa+3x)}{4\kappa^4\theta^2}
\]
and again $A_{11}$ and $A_{22}$ are non-negative provided that $\kappa >0$.
\end{itemize}

 \item[\bf ii) ] $\beta_1 \neq 0$ and $\mu\neq 0$. If the discriminant (\ref{delta13}) is negative, then the minor $d_{2,4}$ is negative and the matrix $A$ cannot be positive semidefinite. Hence we have to assume $\Delta_{2,4}=0$ or $\Delta_{2,4}>0$.
 
 \begin{itemize}
 \item [\bf ii.1) ] $\Delta_{2,4}=0$. Since $\beta_1\neq 0$ and $\lambda,\kappa\neq 0$, from (\ref{delta13}) we get 
\[
\beta_1=-\frac{x\tau}{\kappa\lambda}\alpha_2.
\]
As a consequence $A_{33}=0$ and $d_{1,2}=-A_{34}^2$, $d_{1,4}=-A_{23}^2$ and $d_{2,4}=-A_{13}^2$ so that all the elements $A_{13}$, $A_{23}$ and $A_{43}$ must vanish. These conditions in turn imply
\[\begin{split}
& \kappa\lambda^2=-x\tau\mu\alpha_2\theta,\\
&\kappa^2\lambda^2\alpha_1=\alpha_{2}\left(x^2\tau^2+\kappa(x+\kappa)\lambda\right).
\end{split}\]
Since $\kappa\lambda\neq 0$ and then $x\tau\mu\alpha_2\neq 0$, these equations fix the values of $\alpha_1$ and $\alpha_2$ as follows  
\[
\alpha_1=-\frac{x^2\tau^2+\kappa\lambda\mu}{x\kappa\mu\tau\theta}, \quad \alpha_2=-\frac{\kappa\lambda^2}{x\mu\tau\theta}.
\]
At this point all the terms $A_{i3}$, $i=1..4$ are equal to zero. The remaining elements are given by
\[\begin{split} 
 &  A_{11}= \frac{1}{\mu\theta},\;   A_{12}= \frac{x(x+\kappa)\tau^2+\kappa\lambda(\mu-\kappa)}{2x\kappa\mu\tau\theta},\; A_{14}=\kappa A_{12}, \\
&  A_{22}= -\frac{(x+\kappa) \lambda}{x\mu\theta},\; A_{24}=\kappa A_{22},\; A_{44}= \kappa^2A_{22}.
\end{split}\]
Notice that from $A_{ii}\geq 0$, $i=1..4$, we get 
\beq\label{mux_cond}
\mu>0,\qquad x(x+\kappa)\lambda \leq 0.\eeq 
The 3-by-3 minors are all zero and the non zero 2-by-2 minors are just $d_{3,4}$ and $d_{2,3}=k^2d_{3,4}$. Since
$$
d_{3,4}=-\frac{\kappa^2\lambda^2\mu^2+2\kappa\lambda(x(x+\kappa)\tau^2-\kappa^2\lambda)\mu+(x(x+\kappa)\tau^2+\kappa^2\lambda)^2}{4x^2\tau^2\theta^2\kappa^2\mu^2}.
$$
the necessary and sufficient condition in order to have a positive semidefinite matrix is the following
\begin{equation}\label{dismu1}
\kappa^2\lambda^2\mu^2+2\kappa\lambda(x(x+\kappa)\tau^2-\kappa^2\lambda)\mu+(x(x+\kappa)\tau^2+\kappa^2\lambda)^2\leq 0.
\end{equation}
By virtue of \eqref{mux_cond} its discriminant with respect to $\mu$, say $\Delta_{3,4}$,  is always non-negative,
$$
\Delta_{3,4}=-16x(x+\kappa)\kappa^4 \lambda^3\tau^2\geq0.
$$
Therefore, the quadratic inequality (\ref{dismu1}) is satisfied iff $\mu\in [\mu_2^-, \mu_2^+]$, where 
\begin{equation}\label{mupm2}
\mu_2^{\pm} = \frac{\left(|\tau|\sqrt{-\lambda x(x+\kappa)}\pm |\lambda|\kappa\right)^2}{\kappa\lambda^2}.
\end{equation}
Notice that  $\mu_2^{\pm}>0$ provided that $\kappa>0$. Moreover, the interval collapse to a point when either $\nu=0$ or $\nu=\kappa$, so that it follows $\mu=\kappa$. \\ Summarizing, if $\Delta_{2,4}=0$ thermodynamic consistency is ensured if  $\lambda \nu(\nu-\kappa) \leq 0$, $\kappa>0$, $(\nu-\kappa)\tau\mu\neq 0$ and $\mu \in [\mu_2^-, \mu_2^+]$, where $\mu_2^{\pm}$ are given by (\ref{mupm2}).

\smallskip
 \item [\bf ii.2) ] $\Delta_{2,4}>0$. Since $\Delta_{2,4}>0$, $\beta_1\neq 0$ and the diagonal terms of the matrix $A$ (1-by-1 minors) must be non-negative, the following inequalities hold:
\begin{equation}\label{ineq}
\lambda\beta_1>0, \quad \mu \geq 0, \quad \lambda\kappa\beta_1+x\tau\alpha_2>0, \quad \lambda(\tau\alpha_2-\lambda\beta_1) \geq 0.
\end{equation}
Applying these inequalities, the 2-by-2 minor $d_{2,4}$ given by (\ref{d13}) is non-negative if $\mu$ belongs to the real positive interval $\mu \in [\mu_3^-, \mu_3^+]$, where 
\[
\mu_3^{\pm} = \frac{\left(\sqrt{\theta(\kappa\lambda\beta_1+x\tau\alpha_2)}\pm|\lambda|\right)^2}{\lambda\beta_1\theta}.
\]
We need to check the other minors. The 2-by-2 minors different from zero are $d_{1,4}$, $d_{3,4}$ and $d_{2,3}=k^2d_{3,4}$, $d_{1,2}=k^2d_{1,4}$; the 3-by-3 minors different from zero are $d_4$ and $d_2=\kappa^2 d_4$. 
For future convenience, we introduce the parameter $\phi$ as
\[
\phi=\frac{\alpha_1\alpha_2-\beta_1^2}{\lambda}.
\]
Minors $d_{1,4}$ and $d_4$ are quadratic polynomials  in $\mu$, whereas $d_{3,4}$ does not contain $\mu$. First, let us consider 
\begin{equation}\label{d14}
\begin{split}
d_{1,4}=&-\frac{1}{4\lambda^4}\Big(\lambda^2\alpha_2^2\mu^2-2\lambda\left(\lambda^3\kappa\phi+(\tau\alpha_2-\lambda\beta_1)(\kappa\lambda\beta_1+\tau x \alpha_2)\right)\mu+\\
&+\left(\kappa\lambda(\tau\beta_1-\lambda\alpha_1)+x\tau(\tau\alpha_2-\lambda\beta_1)\right)^2\Big).
\end{split}\end{equation}
The discriminant  of $d_{1,4}$ with respect to $\mu$ is given by
\[
\Delta_{1,4}=\frac{\kappa\phi(\tau\alpha_2-\lambda\beta_1)(\kappa\lambda\beta_1+x\tau\alpha_2)}{\lambda^3}.
\]
$\Delta_{1,4}$ must be non-negative, otherwise the minor $d_{1,4}$ would be negative. By virtue of inequalities (\ref{ineq}),
  $\Delta_{1,4}\ge0$ implies
\begin{equation}\label{ineqk}
\kappa\phi\geq 0.
\end{equation}
We consider now two sub-cases: a) $\alpha_2\neq 0$ and b) $\alpha_2= 0$.

\smallskip
{\bf a)} 
Let us assume $\alpha_2\neq 0$. In this case equation (\ref{d14}) implies that $\mu$ must belong to the interval $[\mu_4^-,\mu_4^+]$ (which can collapse into a single point if $\Delta_{1,4}=0$) where
\[
\mu_4^{\pm} = \frac{\left(\sqrt{\lambda^2\kappa\phi}\pm\sqrt{\displaystyle\frac{(\tau\alpha_2-\lambda\beta_1)(\lambda\kappa\beta_1+x\tau\alpha_2)}{\lambda}}\right)^2}{\alpha_2^2}.
\]

Let us now consider the minor
\[
d_4=-\frac{\beta_1\phi}{4\lambda}\mu^2+2P\mu-Q,
\]
where $P$ and $Q$ are polynomials of the parameters and the temperature,
\begin{equation}\label{AB}\begin{split}
P=&\frac{\phi}{8\lambda^2\theta}\left(x\theta\beta_1\tau^2+x\tau\theta(\alpha_2-\lambda\alpha_1)+\lambda(\lambda+2\kappa\beta_1\theta)\right)+\frac{\beta_1(\tau\alpha_2-\lambda\beta_1)-\alpha_2^2}{8\lambda^2\theta},\\ 
Q=&\frac{\beta_1\left(\alpha_2x\tau^2+\beta_1(k-x)\lambda\tau-\lambda^2\kappa\beta_1\right)^2}{4\lambda^5}+\frac{(\tau\alpha_2-\lambda\beta_1)\left((\lambda\kappa\beta_1+x\tau\alpha_2)\theta-\lambda^2\right)^2}{4\lambda^5\theta^2}\\
&-\frac{((\lambda\kappa\beta_1+x\tau\alpha_2)\theta-\lambda^2)(\alpha_2-\lambda\alpha_1+\tau\beta_1)(\alpha_2x\tau^2+\beta_1(k-x)\lambda\tau-\lambda^2\kappa\beta_1)}{4\lambda^5\theta}.
\end{split}\end{equation}
The discriminant of $d_4$ with respect to $\mu$ can be written as
\begin{equation}\label{parth}
\Delta_4=\frac{d_{3,4}}{4\lambda^2\theta^2}\Big(4\kappa\lambda\beta_1\phi\theta-(\alpha_2-x\tau\phi\theta)^2\Big).
\end{equation}
where the minor $d_{3,4}$ takes the following form,
\[
d_{3,4}=\phi-\frac{\left(\lambda\alpha_1-\tau\beta_1+\alpha_2\right)^2}{4\lambda^2}.
\]
This expression excludes the possibility $\phi<0$. The equality $\phi=0$ must  also be excluded. Indeed,  the vanishing of $\phi$ imply $\alpha_1\alpha_2-\beta_1^2=0$ (by its definition) and gives
$$
d_{3,4}=-\frac{(\lambda\alpha_1^2-\tau\alpha_1\beta_1+\beta_1^2)^2}{4\lambda^2\alpha_1^2}.
$$
The numerator is forced to vanish and this yields
\[
\tau=\frac{\lambda\alpha_1^2+\beta_1^2}{\alpha_1\beta_1}. 
\]
After replacing this expression for $\tau$ and $\phi=0$, we obtain $d_4=-{\beta_1^3}/{4\lambda\alpha_1^2\theta^2}$ and  $A_{11}={\beta_1}/{\lambda}$; since $\beta_1\neq 0$, the requirements $d_4\geq 0$ and $A_{11}\geq 0$ are not compatible.
We then conclude that $\phi$ must be positive. Consequently,
because of  \eqref{ineq}$_1$, the coefficient of $\mu^2$ in the expression of $d_{4}$ is negative.   Then the discriminant $\Delta_4$ must be non-negative, otherwise the minor $d_4$  would be negative. 

-- If $d_{3,4} > 0$, the factor in round brackets  on the right side of (\ref{parth}) must be non-negative, namely
\begin{equation}\label{kappaeq}
 \kappa\geq \frac{(\alpha_2-x\tau\phi\theta)^2}{4\lambda\beta_1\phi\theta}. 
\end{equation}
Notice that the right hand side of this inequality is non negative due to the constraints on the parameters (\ref{ineq}), (\ref{ineqk}) and $\phi>0$. Also, if (\ref{kappaeq}) is verified, it is easy to show that inequality (\ref{ineq})$_3$ holds. Indeed we have
\begin{equation}\label{condk}
\lambda\kappa\beta_1+x\tau\alpha_2>\frac{(\alpha_2+x\tau\phi\theta)^2}{4\phi\theta}.
\end{equation}
Inequality (\ref{kappaeq}) involves $x=\nu-\kappa$, so it results in a restriction of the admissible values either of $\nu$, if $\tau\neq 0$, or of $\kappa$, if $\tau=0$. When $\tau\neq 0$ we get
\begin{equation}\label{inu2}
\nu \in I_{\nu}^2= [\nu^-,\nu^+], \;\;\; \nu^{\pm}=\kappa+\frac{\alpha_2\pm 2\sqrt{\kappa\lambda\beta_1\phi\theta}}{\tau\phi\theta}.
\end{equation}
If $\tau=0$ we get 
$
\kappa\geq   {\alpha_2^2}/{4\lambda\beta_1\phi\theta}. 
$
The inequality  $d_{3,4}> 0$ can be viewed as a relation determining a range for the admissible values of $\phi$ (or, alternatively, of $\alpha_1$, since we are now considering the case $\alpha_2\neq 0$). Indeed, by making explicit the dependence on $\phi$, we have 
\[
d_{3,4}=-\frac{1}{4\alpha_2^2}\left(\lambda^2\phi^2+2\left(\beta_1(\tau\alpha_2-\lambda\beta_1)+\alpha_2^2\right)\phi-\frac{\left(\beta_1(\tau\alpha_2-\lambda\beta_1)-\alpha_2^2\right)^2}{\lambda^2}\right)
\]
The discriminant of $d_{3,4}$ with respect to $\phi$ is non negative, again due to the constraints (\ref{ineq}). Indeed we have
\[
\Delta_{3,4}=\frac{\beta_1(\tau\alpha_2-\lambda\beta_1)}{\alpha_2^2}.
\]
From $d_{3,4} >0$ then it follows that $\phi$ must be in the interval $(\phi_1^-, \phi_1^+)$ where
\begin{equation}\label{phiint}
\phi_1^{\pm}=\frac{\left(\alpha_2\pm\sqrt{\beta_1(\tau\alpha_2-\lambda\beta_1)}\right)^2}{\lambda^2}
\end{equation}
Notice that this range is compatible with $\phi>0$. In terms of $\alpha_1$, we get
\begin{equation}\label{phiint1}
\alpha_1\in I_{\alpha_1}=[\alpha_1^-, \alpha_1^+], \;\;\; \alpha_1^{\pm}=\frac{\tau\beta_1+\alpha_2\pm\sqrt{\beta_1(\tau\alpha_2-\lambda\beta_1)}}{\lambda}
\end{equation}
If (\ref{phiint1}) and (\ref{kappaeq}) are satisfied, then $d_4\geq 0$ implies
\[
\mu \in I_{\mu}^5 = \left[4\lambda\frac{P-\sqrt{\Delta_4}}{\beta_1\phi},4\lambda\frac{P+\sqrt{\Delta_4}}{\beta_1\phi}\right]
\]

-- It remains to consider the possibility $d_{3,4}=0$. If this is the case, the discriminant (\ref{parth}) vanishes and the interval $I_{\mu}^5$ collapses to a single point $\mu=4\lambda(P-\sqrt{\Delta_4})/{\beta_1\phi}$,
where you must enter the value of $\phi$ that solves $d_{3,4}=0$, i.e.
\begin{equation}\label{phiequ}
\phi=\frac{\left(\lambda\alpha_1-\tau\beta_1+\alpha_2\right)^2}{4\lambda^2}.
\end{equation}
By replacing $\phi$ in terms of the other variables, i.e. $\phi=(\alpha_1\alpha_2-\beta_1^2)/\lambda$, equation (\ref{phiequ})  can be written as a quadratic equation to determine $\alpha_1$, 
\[
\alpha_1^2-2\frac{\tau\beta_1+\alpha_2}{\lambda}\alpha_1+\frac{4\lambda\beta_1^2+(\tau\beta_1-\alpha_2)^2}{\lambda^2}=0.
\]
Notice that this equation has always at least a real solution, since its discriminant is non negative. In particular, $\alpha_1=0$ is a solution provided that
$$
4\lambda\beta_1^2+(\tau\beta_1-\alpha_2)^2=0.
$$
Since the discriminant $\Delta_{4}$ given by (\ref{parth}) vanishes, condition (\ref{kappaeq}) on $\kappa$ does not apply 
and inequality (\ref{ineq})$_3$ is no more automatically verified, but can be seen as a relation identifying the possible values of $x$, i.e. of $\nu$. All the other conditions remain unchanged as in the previous case $d_{3,4}>0$.

\smallskip
{\bf b)} 
Finally, to complete the analysis, we must go back to equation (\ref{d14}) for $d_{1,4}$ and consider the case $\alpha_2=0$. From the 1-by-1 minors we get the inequalities 
\begin{equation}\label{ineqalpha}
\beta_1<0,\;  \lambda <0, \; \kappa >0. 
\end{equation}
Also, $\mu$ must be in the positive set $I_3^{\mu}$. The minor $d_{1,4}$ gives another interval for $\mu$, i.e. $\mu \in I_6^{\mu} = [\mu_6, \infty)$, where 
\[
\mu_6=-\frac{\left(\kappa(\lambda\alpha_1-\tau\beta_1)+x\tau\beta_1\right)^2}{4\kappa\lambda\beta_1^2} 
\]
The only minors to consider are now $d_4$ and $d_{3,4}$. The minor $d_4$ is given by
\begin{equation}\label{d4p1}
d_4=\frac{\beta_1^3}{4\lambda^2}\mu^2+2\hat{P}\mu-\hat{Q}
\end{equation}
where $\hat{P}$ and $\hat{Q}$ are $P$ and $Q$ as given in (\ref{AB}) evaluated at $\alpha_2=0$. By \eqref{ineqalpha}$_1$ the coefficient of $\mu^2$ in (\ref{d4p1}) is negative. Hence $d_4\ge0$ provided that its discriminant 
\begin{equation}\label{parthi}
\Delta_4=-\frac{d_{3,4}\beta_1^3}{4\lambda^4\theta}\Big(\tau^2x^2\beta_1\theta+4\kappa\lambda^2\Big).
\end{equation}
is non negative. 

\smallskip
If $d_{3,4} >0$, then  (\ref{parthi}) requires
$$
\tau^2x^2\beta_1\theta+4\kappa\lambda^2\geq 0,
$$
so that $\nu $ must belong to the interval $ I_{\nu}^2$ as given by (\ref{inu2}) and evaluated in $\alpha_2=0$ if $\tau\neq 0$,  whereas it is automatically satisfied if $\tau=0$.
If $\alpha_2=0$ from (\ref{phiint}) it follows $\phi=-\beta_1^2/\lambda$. Actually, after replacing this expression for $\phi$ and $\alpha_2=0$, $d_{3,4}$ can be represented as a polynomial in $\alpha_1$,
\beq\label{d34_spec}
d_{3,4}=-\frac{\alpha_1^2}{4}+\frac{\tau\beta_1\alpha_1}{2\lambda}-\frac{(\tau^2+4\lambda)\beta_1^2}{4\lambda^2}.
\eeq
If $\alpha_1\neq 0$, $d_{3,4}>0$ yields an interval for $\alpha_1$, i.e. $\alpha_1 \in I^{\alpha_1}_{1}=(\alpha_1^-, \alpha_1^+)$, where 
$$
\alpha_1^{\pm}={(\tau \pm 2\sqrt{|\lambda|})\beta_1}/{\lambda}.
$$
If $\alpha_1=0$, $d_{3,4}>0$ gives  $\lambda< -{\tau^2}/{4}$. In both cases, the relation $d_4\geq 0$ defines an interval for $\mu$, corresponding to the interval $I_5^{\mu}$ evaluated at $\alpha_2=0$.

\smallskip
If $d_{3,4}=0$, (\ref{parthi}) implies that   $\Delta_4$ vanishes and $\mu=-4\lambda^2(\hat{P}-\sqrt{\Delta_4})/{\beta_1^3}$, where we have to replace $\alpha_1$ with a solution of \eqref{d34_spec}. In particular, $\alpha_1=0$ is a solution provided that $\lambda=-{\tau^2}/{4}$.
Otherwise equation \eqref{d34_spec} has always at least one non trivial solution, since its discriminant  is non negative. 

\end{itemize}
\end{itemize}
\newpage
The following tables summarize the results of all items.

\bigskip\hskip-1cm
\small
\begin{table}[h]
\caption{}
\begin{tabular}{| >{\centering}p{.5cm} |>{\centering}p{2.5cm} |>{\centering}p{3cm} |>{\centering}p{1.5cm} |>{\centering}p{2.7cm} | >{\centering\arraybackslash}p{3cm}|}
\hline
\multicolumn{6}{|c|}{\textbf{Summary table for the LSO consistency conditions on material parameters}}\\
\hline
& $\boldsymbol{\mu}$ & $\boldsymbol{\nu}$& $\boldsymbol{\kappa}$& $\boldsymbol{\tau}$& $\boldsymbol{\lambda}$\\
\hline
1& $=\kappa$ & $=\kappa$ & $>0$ & $\neq 0$ & - \\
\hline
2 &  $=\kappa$ & $\geq 0$, $\neq \kappa$ & $>0$ & $\neq 0$ & -\\
\hline
3& $\in I_1^{\mu}$ &  $\lambda(\nu-\kappa)>0$ & $>0$ & $\neq 0$ &-\\
\hline
4 & $0$ & $\frac{\kappa(\lambda+\tau^2)}{\tau^2}$ & $>0$ & $\neq 0$ &-\\
\hline
5& $0$ & - & $>0$ & - & $\leq \frac{\tau^2\nu}{\kappa}, \neq -\frac{\nu(\nu-\kappa)\tau^2}{\kappa^2}$ \\
\hline
6& $0$ &\begin{small} $\neq 0, \neq \kappa$, $(\nu-\kappa)\tau>0$ \end{small}& $>0$ & $\neq 0$, $(\nu-\kappa)\tau> 0$ & $-\frac{\nu(\nu-\kappa)\tau^2}{\kappa^2}$ \\
\hline
7& $0$ & \begin{small}$\neq 0, \neq \kappa$, $(\nu-\kappa)\tau<0$\end{small} & $>0$ & $\neq 0$, $(\nu-\kappa)\tau< 0$ & $-\frac{\nu(\nu-\kappa)\tau^2}{\kappa^2}$ \\
\hline
8 & $\in I_2^{\mu}$ & $\lambda\nu(\kappa-\nu)\geq 0, \nu\neq \kappa$ & $>0$ & $\neq 0$ & -\\
\hline
9 & $\in I_3^{\mu}\cap I_{4}^{\mu}\cap I_5^{\mu}$ & $\in I_{\nu}^1$& $> 0$ & $\lambda(\tau\alpha_2-\lambda\beta_1)\geq 0$, $\neq 0$ & $\lambda\beta_1>0$\\
\hline
10 & $\in I_3^{\mu}\cap I_{4}^{\mu}\cap I_5^{\mu}$ & - & $\geq \frac{\alpha_2^2}{4\lambda\beta_1\phi\theta}$ & $ 0$ & $<0$\\
\hline
11 & $\in I_3^{\mu}\cap I_{4}^{\mu}\cap I_5^{\mu}$ & $\lambda\kappa\beta_1+(\nu-\kappa)\tau\alpha_2>0$ & $>0$ & $\lambda(\tau\alpha_2-\lambda\beta_1)\geq 0$ & $\lambda\beta_1>0$\\
\hline
12 & $\in I_3^{\mu}\cap I_{4}^{\mu}\cap I_5^{\mu}$ & $\lambda\kappa\beta_1+(\nu-\kappa)\tau\alpha_2>0$ & $>0$ & $\tau\alpha_2-\lambda\beta_1\leq 0$ & $-\frac{(\beta_1\tau-\alpha_2)^2}{4\beta_1^2}$\\
\hline
13 & $\in I_3^{\mu}\cap I_{5}^{\mu}\cap I_6^{\mu}$ & $\in I_{\nu}^2$ & $>0$ & - & $<0$\\
\hline
14 & $\in I_3^{\mu}\cap I_{5}^{\mu}\cap I_6^{\mu}$ & $\in I_{\nu}^2$ & $>0$ & - & $<-\frac{\tau^2}{4} $\\
\hline
15 & $\in I_3^{\mu}\cap I_{5}^{\mu}\cap I_6^{\mu}$ & - & $>0$ & - & $< 0$\\
\hline
16 & $\in I_3^{\mu}\cap I_{5}^{\mu}\cap I_6^{\mu}$ & - & $>0$ & - & $-\frac{\tau^2}{4}$\\
\hline
\end{tabular}
\label{tab1}
\end{table}

\newpage
\rotatebox{90}{
\begin{tabular}{| >{\centering}p{0.5cm} | >{\centering}p{2.5cm} |>{\centering}p{2.cm} |>{\centering}p{2.5cm} |>{\centering}p{3.5cm} |>{\centering}p{2.5cm} | >{\centering\arraybackslash}p{2.5cm}|}
\hline
\multicolumn{7}{|c|}{\textbf{Summary table of the coefficients appearing in the quadratic free energy $\psi$ given by (\ref{free_en})}}\\
\hline
& $\boldsymbol{\beta_1}$ & $\boldsymbol{\beta_2}$& $\boldsymbol{\beta_3}$& $\boldsymbol{\alpha_1}$& $\boldsymbol{\alpha_2}$& $\boldsymbol{\alpha_3}$\\
\hline
1& $\frac{\tau}{\lambda}\alpha_2$ & $\frac{\nu\tau^2+\kappa\lambda}{\lambda^2}\alpha_2$ & $\frac{\tau\nu}{\lambda}\alpha_2$ & $\frac{\tau^2+\lambda}{\lambda^2}\alpha_2$ & s.t. $\tau\alpha_2\geq \frac{\lambda^2}{4\kappa\theta}$ & $\frac{\lambda\kappa^2+\tau^2\nu^2}{\lambda^2}\alpha_2$\\
\hline
2 & $\frac{\tau}{\lambda}\alpha_2$ & $\frac{\nu\tau^2+\kappa\lambda}{\lambda^2}\alpha_2$ & $\frac{\tau\nu}{\lambda}\alpha_2$ & $\frac{\tau^2+\lambda}{\lambda^2}\alpha_2$ & $\in (\alpha_2^-, \alpha_2^+)$ & $\frac{\lambda\kappa^2+\tau^2\nu^2}{\lambda^2}\alpha_2$ \\
\hline
3&$0$ & $\frac{\kappa \lambda}{\tau(\nu-\kappa)\theta}$ & $\frac{\lambda}{\theta}$ & $\frac{\lambda}{\tau(\nu-\kappa)\theta}$ & $\frac{\lambda^2}{\tau(\nu-\kappa)\theta}$ & $\frac{\kappa^2\lambda+\tau^2(\nu-\kappa)^2}{\tau(\nu-\kappa)\theta}$ \\
\hline
4 & $0$ & $\frac{\tau}{\theta}$ & $\frac{\lambda}{\theta}$ & $\frac{\tau}{\kappa\theta}$ & $\frac{\lambda\tau}{\kappa\theta}$ & $\frac{\kappa(\lambda+\tau^2)}{\tau\theta}$\\
\hline
5&$\frac{\lambda^2-\tau\alpha_2(\nu-\kappa)\theta}{\kappa\lambda\theta}$ & $\frac{\tau}{\theta}$ & $\frac{\lambda}{\theta}$ & $\frac{\left(\tau\alpha_2(\nu-\kappa)^2\theta-(\nu-2\kappa)\lambda^2\right)\tau}{\kappa^2\lambda^2\theta}$ & $\in I_{\alpha_2}$ & $\frac{\tau\nu}{\theta}$ \\
\hline
6&$\frac{\lambda^2-\tau\alpha_2(\nu-\kappa)\theta}{\kappa\lambda\theta}$ & $\frac{\tau}{\theta}$ & $\frac{\lambda}{\theta}$ & $\frac{\left(\tau\alpha_2(\nu-\kappa)^2\theta-(\nu-2\kappa)\lambda^2\right)\tau}{\kappa^2\lambda^2\theta}$ & $\leq \frac{(\nu-\kappa)^2(3\nu+\kappa)\tau^3}{4\kappa^4\theta}$ & $\frac{\tau\nu}{\theta}$ \\
\hline
7&$\frac{\lambda^2-\tau\alpha_2(\nu-\kappa)\theta}{\kappa\lambda\theta}$ & $\frac{\tau}{\theta}$ & $\frac{\lambda}{\theta}$ & $\frac{\left(\tau\alpha_2(\nu-\kappa)^2\theta-(\nu-2\kappa)\lambda^2\right)\tau}{\kappa^2\lambda^2\theta}$ & $\geq \frac{(\nu-\kappa)^2(3\nu+\kappa)\tau^3}{4\kappa^4\theta}$ & $\frac{\tau\nu}{\theta}$ \\
\hline
8& $\frac{\lambda}{\mu\theta}$ & $\frac{\kappa\lambda}{\tau(\kappa-\nu)\theta}$ & $0$ & $\frac{\tau^2(\kappa-\nu)^2+\kappa\lambda\mu}{\mu\tau\kappa(\kappa-\nu)\theta}$ & $\frac{\kappa\lambda^2}{\mu\tau(\kappa-\nu)\theta}$ & $\frac{\kappa^2\lambda}{\tau(\kappa-\nu)\theta}$ \\
\hline
9 & $\lambda\beta_1>0$ & $\frac{\lambda\kappa\alpha_1+\tau(\nu-\kappa)\beta_1}{\lambda}$ & $\kappa\beta_1+\frac{\tau(\nu-\kappa)\alpha_2}{\lambda}$ & $\in I_{\alpha_1}$ & $\neq 0$ & \begin{small}$\kappa^2\alpha_1+2\kappa\tau\frac{\nu-\kappa}{\lambda}\beta_1+\tau^2\frac{(\nu-\kappa)^2}{\lambda^2}\alpha_2$ \end{small}\\
\hline
10 & $<0$ & $\kappa\alpha_1$ & $\kappa\beta_1$ & $\in I_{\alpha_1}$ & $\neq 0$ & \begin{small}$\kappa^2\alpha_1$ \end{small}\\
\hline
11 & $\lambda\beta_1>0$ & $\frac{\lambda\kappa\alpha_1+\tau(\nu-\kappa)\beta_1}{\lambda}$ & $\kappa\beta_1+\frac{\tau(\nu-\kappa)\alpha_2}{\lambda}$ &\begin{small}$\alpha_1^2-2\frac{\tau\beta_1+\alpha_2}{\lambda}\alpha_1+\frac{4\lambda\beta_1^2+(\tau\beta_1-\alpha_2)^2}{\lambda^2}=0$\end{small} &  $\neq 0$ & \begin{small}$\kappa^2\alpha_1+2\kappa\tau\frac{\nu-\kappa}{\lambda}\beta_1+\tau^2\frac{(\nu-\kappa)^2}{\lambda^2}\alpha_2$ \end{small}\\
\hline
12 & $<0$ & $\frac{\tau(\nu-\kappa)\beta_1}{\lambda}$ & $\kappa\beta_1+\frac{\tau(\nu-\kappa)\alpha_2}{\lambda}$ & $0$ &  $\neq 0$ & \begin{small}$2\kappa\tau\frac{\nu-\kappa}{\lambda}\beta_1+\tau^2\frac{(\nu-\kappa)^2}{\lambda^2}\alpha_2$ \end{small}\\
\hline
13 & $<0$ & $\frac{\lambda\kappa\alpha_1+\tau(\nu-\kappa)\beta_1}{\lambda}$ & $\kappa\beta_1$ &\begin{small}$\in I_{\alpha_1}^2$\end{small} &  $ 0$ & \begin{small}$\kappa^2\alpha_1+2\kappa\tau\frac{\nu-\kappa}{\lambda}\beta_1$ \end{small}\\
\hline
14 & $<0$ & $\frac{\tau(\nu-\kappa)\beta_1}{\lambda}$ & $\kappa\beta_1$ &\begin{small}$0$\end{small} &  $ 0$ & \begin{small}$2\kappa\tau\frac{\nu-\kappa}{\lambda}\beta_1$ \end{small}\\
\hline
15 & $<0$ & $\frac{\tau(\nu-\kappa)\beta_1}{\lambda}$ & $\kappa\beta_1$ &\begin{small}$\alpha_1^2-2\frac{\tau\beta_1}{\lambda}\alpha_1+\frac{4\lambda\beta_1^2+(\tau\beta_1)^2}{\lambda^2}=0$\end{small} &  $ 0$ & \begin{small}$2\kappa\tau\frac{\nu-\kappa}{\lambda}\beta_1$ \end{small}\\
\hline
16 & $<0$ & $\frac{\tau(\nu-\kappa)\beta_1}{\lambda}$ & $\kappa\beta_1$ &\begin{small}$0$\end{small} &  $ 0$ & \begin{small}$2\kappa\tau\frac{\nu-\kappa}{\lambda}\beta_1$ \end{small}\\
\hline
\multicolumn{7}{|c|}{\bf The values of the other coefficients are as follows: $\beta_4 =\kappa\beta_1, \qquad\beta_5 =\kappa\alpha_2 , \qquad \beta_6=\kappa\beta_3,\qquad \alpha_4=\kappa\beta_5$}\\
\hline
\end{tabular}}

\normalsize

\end{document}